\documentclass{aa}  

\usepackage{graphicx}
\usepackage{txfonts}

\usepackage{newtxtext,newtxmath}
\usepackage[T1]{fontenc}
\usepackage{graphicx}	% Including figure files
\usepackage{amsmath}	% Advanced maths commands

\usepackage{xcolor}

\newcommand{\NII}{[N\,{\sc II}]}
\newcommand{\SII}{[S\,{\sc II}]}
\newcommand{\OIII}{[O\,{\sc III}]}

\newcommand{\Ha}{H$\alpha$}
\newcommand{\Hb}{H$\beta$}

\begin{document}

   \title{The Fornax3D project: The environmental impact on gas metallicity gradients in Fornax cluster galaxies}
   \titlerunning{The environmental impact in Fornax cluster galaxies}

%   \subtitle{I. Overviewing the $\kappa$-mechanism}

   \author{M. A. Lara-L\'opez \inst{1}
          \and
       P. M. Gal\'an-de Anta\inst{1,2}
          \and
        M. Sarzi\inst{1}
        \and
        E. Iodice\inst{3,4}
                \and{T. A. Davis}\inst{5}
                \and{N. Zabel}\inst{6}
 \and{E. M. Corsini}\inst{7,8}
\and{ P. T. de Zeeuw}\inst{9,10}
       \and{K. Fahrion}\inst{4,11}
        \and{J. Falc\'on-Barroso}\inst{12,13}
        \and{D. A. Gadotti}\inst{4}
        \and{R. M. McDermid}\inst{14,15}
        \and{F. Pinna}\inst{16}
        \and{V. Rodriguez-Gomez}\inst{17}
        \and{G. van de Ven}\inst{18}
              \and{L. Zhu }\inst{19}
              \and{L. Coccato }\inst{4}
              \and{M. Lyubenova}\inst{4}
      \and{I. Mart\'in-Navarro}\inst{12,13}
                  }

   \institute{Armagh Observatory and Planetarium, College Hill, Armagh, BT61 DG, UK
              \email{Maritza.Lara-Lopez@armagh.ac.uk}
         \and
             Astrophysics Research centre, School of Mathematics and Physics, Queen’s University Belfast, Belfast BT7 INN, UK
             \and
              INAF - Osservatorio Astronomico di Capodimonte, Salita Moiariello 16, 80131, Napoli, Italy
              \and
European Southern Observatory, Karl-Schwarzschild-Straße 2, 85748 Garching bei München, Germany
                    \and                
Cardiff Hub for Astrophysics Research \&\ Technology, School of Physics \&\ Astronomy, Cardiff University, Queens Buildings, The Parade, Cardiff, CF24 3AA, UK
              \and
Kapteyn Astronomical Institute, University of Groningen PO Box 800, 9700 AV Groningen, The Netherlands
              \and
 Dipartimento di Fisica e Astronomia ‘G. Galilei’, Universit\`{a} di Padova, vicolo dell’Osservatorio 3, 35122 Padova, Italy
              \and
INAF–Osservatorio Astronomico di Padova, vicolo dell’Osservatorio 5, 35122 Padova, Italy
              \and
Sterrewacht Leiden, Leiden University, Postbus 9513, 2300 RA Leiden, The Netherlands
              \and
Max-Planck-Institut für extraterrestrische Physik, Giessenbachstraße, 85741, Garching bei M\"unchen, Germany
              \and
   European Space Agency, European Space Exploration and Research Centre, Keplerlaan 1, 2201 AZ Noordwijk, The Netherlands
             \and
Instituto de Astrof\'isica de Canarias, Vía L\'actea s/n, 38205 La Laguna, Tenerife, Spain
              \and
 Depto. Astrof\'isica, Universidad de La Laguna, Calle Astrof\'isico Francisco S\'anchez s/n, 38206 La Laguna, Tenerife, Spain
              \and
Department of Physics and Astronomy, Macquarie University, Sydney, NSW 2109, Australia
              \and
 ARC Centre of Excellence for All Sky Astrophysics in 3 Dimensions (ASTRO 3D), Australia
  	\and                    
Max-Planck-Institut f\"{u}r Astronomie, K\"{o}nigstuhl 17, 69117, Heidelberg, Germany
              \and
Instituto de Radioastronom\'ia y Astrof\'isica, Universidad Nacional Aut\'onoma de M\'exico, Apdo. Postal 72-3, 58089 Morelia, M\'exico
              \and
Department of Astrophysics, University of Vienna, T\"{u}rkenschanzstrasse 17, 1180 Vienna, Austria
              \and
Shanghai Astronomical Observatory, Chinese Academy of Sciences, 80 Nandan Road, Shanghai 200030, China
             }

   \date{Received TBD; accepted TBD}

% \abstract{}{}{}{}{} 
% 5 {} token are mandatory
 
  \abstract
 %  {TBD}
%   {TBD}
%   {TBD}
   {The role played by environment in galaxy evolution is a current debate in astronomy. The degree to which environment can alter, re-shape, or drive galaxy evolution is a topic of discussion in both fronts, observations and simulations. However, our knowledge of the effect of environment on gas metallicity gradients so far is  limited.
   This paper analyses the gas metallicity gradients for a sample of 10 Fornax cluster galaxies observed with MUSE as part of the Fornax3D project. Detailed maps of emission lines allowed a precise determination of gas metallicity and metallicity gradients. 
   { The integrated gas metallicity of our Fornax cluster galaxies show  slightly higher metallicities ($\sim$0.045 dex) in comparison to a control sample}. In addition, we find signs of a mass and metallicity segregation from the center to the outskirts of the cluster.
    By comparing our Fornax cluster metallicity gradients with a control sample we find a general median offset of $\sim$0.04 dex/R$_{\rm e}$, with 8 of our galaxies showing flatter or more positive gradients. 
    { The intermediate infallers in our Fornax sample show more positive gradients with respect to the control sample.}
    We find no systematic difference  between the gradients of recent and intermediate infallers when considering the projected distance of each galaxy to the cluster center. 
  %Or, "To elucidate"
    To identify the origin of the observed offset in the metallicity gradients, we perform a similar analysis with  data from the TNG50 simulation. We identify 12 subhalos in Fornax-like clusters and compare their metallicity gradients with a control sample of field subhalos. This exercise also shows  a flattening in the metallicity gradients for galaxies in Fornax-like halos, with a median offset of $\sim$0.05  dex/R$_{\rm e}$.  We also analyse the merger history, Mach numbers ($\mathcal{M}$), and ram pressure stripping of our TNG50 sample. We conclude that the observed flattening in metallicity gradients is likely due to a combination of galaxies traveling at supersonic velocities ($\mathcal{M}$$>$1) that are experiencing high ram pressure stripping and flybys.
% In this paper we analyzed the gas metallicity gradients of ELGs in the Fornax cluster. By comparing Fornax-ELGs gradients with galaxies from a control sample we find a systematic offset, with 8 of our galaxies showing flatter gradients, and a general systematic difference of $\sim$0.04 for the whole sample.   
   }


   \maketitle
%
%-------------------------------------------------------------------

\section{Introduction}

The gas metallicity is an important physical parameter directly linked to the star forming history of the host galaxy. Indeed, it provides important clues on the physical properties of the interstellar medium (ISM), and on the history, evolution, formation and growth of galaxies.

For several decades, a great amount of effort was put into the analysis of scaling relations such as the mass-metallicity (M-Z) relation \citep[e.g., ][]{Lequeux79,Tremonti2004,Kewley08,Lara13}. The M-Z relation provides essential insight into galaxy evolution \citep[e.g.,][]{Savaglio05,Erb06,Lara09a,Lara09b,Lara10a,Zahid12,Sanders21,Bellstedt21}, it is sensitive to metal loss due to stellar winds \citep{Spitoni10,Tremonti2004}, supernovae \citep{Brooks07}, active galactic nuclei (AGN) feedback \citep{Lara19}, and environment \citep[][]{Mouhcine07,Cooper08,Ellison09,Garduno21}. 

With the advent of Integral Field Unit (IFU) spectroscopy, extensive analysis of the resolved physical conditions of the ISM within galaxies has been performed as part of large surveys, such as the Sydney-AAO Multi-object Integral-field unit survey \citep[SAMI, ][]{Poetrodjojo18}, the Calar Alto Legacy Integral Field Area survey \citep[CALIFA, ][]{sanchez-Menguiano16} and the Mapping Nearby Galaxies at APO \citep[MANGA, ][]{Zhang17}; and with surveys of nearby galaxies with a angular resolved resolution such as  VIRUS-P Exploration of Nearby Galaxies \citep[VENGA, ][]{Blanc09, Kaplan16}, the Physics at High Angular resolution in Nearby Galaxies \citep[PHANGS, ][]{Kreckel19}, and the MUSE Atlas of Disks \citep[MAD, ][]{Erroz19}.

The spatial information provided by IFU surveys allows us to study in detail resolved maps of gas metallicity for emission line galaxies. Negative gas metallicity gradients are widely found in late-type galaxies in the local universe \citep[e.g., ][]{Zaritsky94,Pilyugin04,Moustakas10,Rupke10a, Pilyugin14,Ho15,sanchez-Menguiano16,Belfiore17,Smen18}. Such negative gradients are consistent with an inside-out growth scenario of the discs \citep[e.g.][]{White91, Mo98, Perez13}.

Recently, a trend of the resolved metallicity gradient with the total stellar mass of a galaxy was observed by \citet{Belfiore17} using a sample of 550 nearby galaxies from MANGA. They find that galaxies with  log(M$_\star$/M$_{\sun}$) $<9.5$ show flatter gradients that steepens for more massive galaxies until  log(M$_\star$/M$_{\sun}$) $\sim$ 10.5, and then flattens slightly again for more massive systems. Similar results are found for SAMI galaxies \citep[][]{Poetrodjojo21}. 
%Low mass galaxies with flatter gradients is a common result found in several samples
However, so far simulations have been unable to reproduce this trend with stellar mass, with more work needed  \citep[e.g.,][]{Hemler21}.

%Variations in gas metallicity can be indicative of events such as gas inflows of pristine or enriched gas, mergers, past bursts of star formation, and interactions.  
Gas metallicity gradients have been shown to be sensitive to processes such as secular evolution and radial migration \citep[e.g.,][]{Friedli94,Vilchez96,Marino16}.  Morphological studies also claim that barred spirals exhibit flatter metallicity gradients than unbarred galaxies \citep[e.g.][]{Vila92, Zaritsky94,Henry99,Kreckel19,Lara21}.

 However, the study of the effect of environment on gas metallicity gradients so far is rather limited, even though observationally there is evidence that between 50-70$\%$ of the galaxy population is in groups and clusters \citep[e.g.,][]{Eke05}. This naturally implies that processes taking place in the group environment can have a significant impact on the evolution of the galaxy population as a whole. Groups and clusters of galaxies have long been considered perfect laboratories to study the effect of feedback processes in galaxies and their role in (re-)shaping galaxy properties and evolution.  For instance, galaxy interactions and mergers can cause gas inflows, drive morphological transformations, trigger star formation, and even lead to activity in the galactic nucleus \citep[e.g.,][]{Lambas03,Nikolic04,Woods2007, Ellison08, davies15, gordon18, Ellison19, Pan19, Shah20}.  However, the extent of the role played by the environment, also known as nature versus nurture, has been a matter of debate for decades \citep[e.g.,][]{DiMatteo05, Hopkins06,Lani13,Paulino19,Tortora20}. 

Galaxies in clusters tend to have higher metallicities by up to $\sim$0.04 dex when compared to  control samples \citep[e.g.,][]{Ellison09,Scudder12}. 
 These higher metallicities can be explained by several scenarios. For instance, if galaxies undergo a burst of star formation as they enter the cluster environment, this may cause the increase of galaxy metallicity  \citep[e.g.,][]{Finlator08}. In addition, strangulation/starvation may be at play, where the gas reservoir of a galaxy is stripped away or truncated due to interactions with the intra-cluster medium (ICM) or other galaxies. During this process, the cessation of pristine gas accretion on to the galaxy prevents the dilution of metals in the ISM \citep[e.g.,][]{Larson80,Bekki02,Maier21}. Alternatively, cluster galaxies that experience inflows of pre-enriched gas, also known as "chemical pre-processing" would also explain higher metallicities \citep[e.g., ][]{peng10,Gupta18}. 

%Indeed, environmental processes that trigger short-lived bursts of star formation may cause a significant, but transient, change in a galaxy's metallicity before it returns to an equilibrium metallicity \citep[e.g.,][]{Finlator08}.  

Recent studies find that environment might be responsible for flattening metallicity gradients. For instance, using galaxies from the  GAs Stripping Phenomena in galaxies with MUSE \citep[GASP survey,][]{Poggianti17} and MANGA,  \citet{Franchetto21} observed  flatter gradients for galaxies in clusters.
Also,  \citet{Kewley10} found that galaxies in pairs show flatter  metallicity gradients. The physical explanation behind the observed flattening however, is still yet to be understood.  \citet{Kewley10} attributed this flattening to large inflows of gas induced by the tidal effects of galaxy interactions,  whereas \citet{Franchetto21}  suggest that cluster galaxies with flatter gradients might have fallen into the cluster sooner and hence experienced environmental effects for a longer time.

%A possible process known as flyby interactions have been largely neglected until recently, and could be responsible for flattening the metallicity gradients. Flybys are rapid, transient events that occur when two independent galaxy halos interpenetrate but detach at a later time \citep[e.g., ][]{Moore96, Sinha12, An19}. The importance of flybys in galaxy evolution has been recently acknowledged since multiple interactions with two or more neighbours are on average flyby-dominated. According to \citet{An19}, flybys substantially outnumber mergers toward z = 0 (by a factor of five) and the multiple interactions are flyby-dominated, hence flyby’s contribution to galactic evolution is stronger than thought.

%and environmental effects seem to have the effect of flattening these profiles \citep[e.g.,][]{Friedli94,Vilchez96,Marino16}.
%The emission lines produced by ionized gas around stars is key to decipher how galaxies evolve. 

%Metallicity gradients \citet{Ho15,sanchez-Menguiano16}

%Strongly depleted in nearby clusters, spiral galaxies make up about 50\% of the galaxies in clusters at redshift z -0.5 (Oemler, Dressler, \& Butcher 1997; Ellis et al. 1997; Smail et al. 1997)

%Metallicity gradients by \citet{Ho15}

%Metallicity gradients for Illustris TNG-50 by \citet{Hemler21}

In this paper we aim to provide insights into the effect of cluster environment on gas metallicity gradients by analyzing a sample of 10 galaxies in the Fornax cluster observed with MUSE.  Additionally, we use TNG50 simulations of Fornax-like subhalos to further investigate the cluster's impact on the galaxies. All together,  we aim to shed light on specific cluster effects and their impact on gas metallicity  gradients.

This paper is structured as follows, in \S \ref{sec:obs} we present a descriptions of the observations and estimation of spectroscopic emission lines. In \S \ref{sec:SampleCharacterization} we characterize the sample in terms of the M-Z relation and mass segregation. Our results are presented in \S \ref{sec:Results}, and our findings are discussed  in  \S \ref{sec:Discussion}. Finally, our conclusions are given in \S \ref{sec:Conclusions}.

%--------------------------------------------------------------------
\section{Observations}\label{sec:obs}

This paper is based on the Fornax3D survey   \citep[F3D,][]{Sarzi18}, which observed  galaxies  brighter than m$_{\rm B}$ $\leq$ 15 mag  in the Fornax cluster within the viral radius with MUSE. The F3D sample consists of 33 galaxies selected from the catalog of the Fornax cluster members by \citet{Ferguson89}. This study focuses on the gas metallicity gradients and hence we restrict our sample to emission line galaxies (ELGs), since these are needed to estimate gas metallicities.  The final sample analyzed consists of 10 late-type and dwarf irregular galaxies.  Nine of the ELGs were observed with a single central pointing (providing the full coverage for seven of them, and a partial coverage for FCC 285 and FCC 290), while FCC 312 was observed with three pointings (see Fig. \ref{fig:MapsAndBPT}). For further details on the observations refer to \citet{Sarzi18}. 
 In addition, the Fornax cluster galaxies included in this paper were classified as either recent or intermediate infallers by \citet{Iodice19} following the prescription of \citet{Rhee17}. This classification was based on the position of the galaxies in the projected phase-space (PPS) diagram R$_{\rm proj}$/R$_{\rm vir}$ vs. V$_{\rm los}$/$\sigma$$_{\rm los}$, where R$_{\rm proj}$ and R$_{\rm vir}$ are the projected and viral radius, respectively, while V$_{\rm los}$ and $\sigma$$_{\rm los}$ are the line of sight radial velocity, and the cluster velocity dispersion, respectively.

%In addition, the Fornax cluster galaxies use in this paper were classified as either recent or intermediate infallers by \citet{​​​​​​​Iodice219}​​​​​​​.

The data reduction and calibration are described in \citet{Sarzi18}. In short, galaxies were observed with MUSE at ESO Very Large Telescope in Chile  between July 2016 and December 2017 in Wide Field Mode \citep{Bacon10}, with a 1 $\times$ 1 arcmin$^2$ field coverage, a spatial sampling of 0.2 $\times$ 0.2 arcsec$^2$, a wavelength range of 4650-9300 \r{A}, and a spectral sampling of 1.25 \r{A} per pixel. The measured spectral resolution was on average FWHM$_{\rm inst}$=2.8 \r{A}, with little variation ($<$0.2  \r{A}) with wavelength and position over the field of view. The observations were done in good seeing conditions with a median FWHM = 0.88 arcsec.
The data were reduced with the MUSE pipeline version 1.6.2 \citep{Weilbacher16}. The sky subtraction was performed by fitting and subtracting a sky model spectrum on each spaxel of the field of view, with an additional cleaning of the residual sky contamination achieved with the Zurich Atmospheric Purge algorithm \citep[ZAP,][]{Soto16}.

The emission lines were measured using the Gas and Absorption Line Fitting code \citep[GandALF,][]{Sarzi06,Falcon06} in the wavelength range between 4800-6800 \r{A}. The emission line fitting was performed spaxel by spaxel in order to provide high spatial resolution maps and abundances. For this study, we use the following emission lines:  \Hb ,  \OIII\ $\lambda$5007, \Ha , \NII\ $\lambda$6584, and \SII\  $\lambda$$\lambda$6717, 31.

\section{Sample characterization}\label{sec:SampleCharacterization}

%Stellar masses were estimated by \citet{Raj19} through the average $g-i$ colour map for each galaxy, following the procedure of \citet{Iodice19}, and then using the empirical relation from \citet{Taylor11}. 
 
We use the stellar masses and effective radii (R$_{\rm e}$) estimated by  \citet{Raj19} for our sample of  ELGs. These parameters were estimated through a photometric analysis and isophotal fit following the methodology of \citet{Iodice19}. From the isophotal fit, the growth curve analysis was used to derive  R$_{\rm e}$ in several bands for all galaxies. The stellar mass of each galaxy was estimated through the integrated $g-i$ color using the empirical relation from \citet{Taylor11}, which adopts a Chabrier initial mass function (IMF). The stellar masses and R$_{\rm e}$ for our sample are listed in Table \ref{tab:Properties}.

For each galaxy, we selected only the star forming (SF) spaxels using the standard  BPT diagram  \NII\ $\lambda$6584 / \Ha \ vs \OIII\ $\lambda$5007 / \Hb\   \citep[][]{Baldwin81}, with the discrimination of \citet{Kauffmann03}. The middle column of Fig. \ref{fig:MapsAndBPT} shows the map of BPT classification for each galaxy. Spaxels classified as Composite, Seyfert and LINERs were selected following the prescription of \citet{Kewley06}, and are displayed in the same figure. With the exception of FCC 179 and  FCC 290, the rest of the galaxies in our sample are mostly characterised by SF regions. FCC 179 shows Seyfert and LINER regions in the center and outskirts, with a ring of SF regions in between. The center of FCC 290 is full of LINER regions with composite regions scattered throughout the galaxy. Hence, it can be classified as a central LINER, or cLIER \citep{Belfiore16}.

From the plethora of available gas metallicity calibrations, we selected  the one of  \citet{Dopita16} defined as:
%Gas metallicities were estimated by using the calibration of 	

 \begin{multline}\label{Eq.Dopita}
 \begin{split}
{\rm 12+log(O/H)}=8.77 + {\rm y}  \\
{\rm y} = {\rm log(} {\rm \NII} / {\rm \SII} ) + 0.264 \times {\rm log(} {\rm \NII} / {\rm H\alpha} ) 
\end{split}
\end{multline}

Among its advantages, this calibration is independent of reddening, since all the emission lines involved are close in wavelength (within 20\AA). We applied this calibration only to the SF spaxels in each galaxy. The obtained gas metallicity maps are shown in Figures \ref{fig:MapsAndBPT}-\ref{fig:MapsAndBPTc}.

\subsection{The M-Z relation for Fornax galaxies}\label{subsec:MZ}

\begin{figure}
	\includegraphics[width=0.5\textwidth]{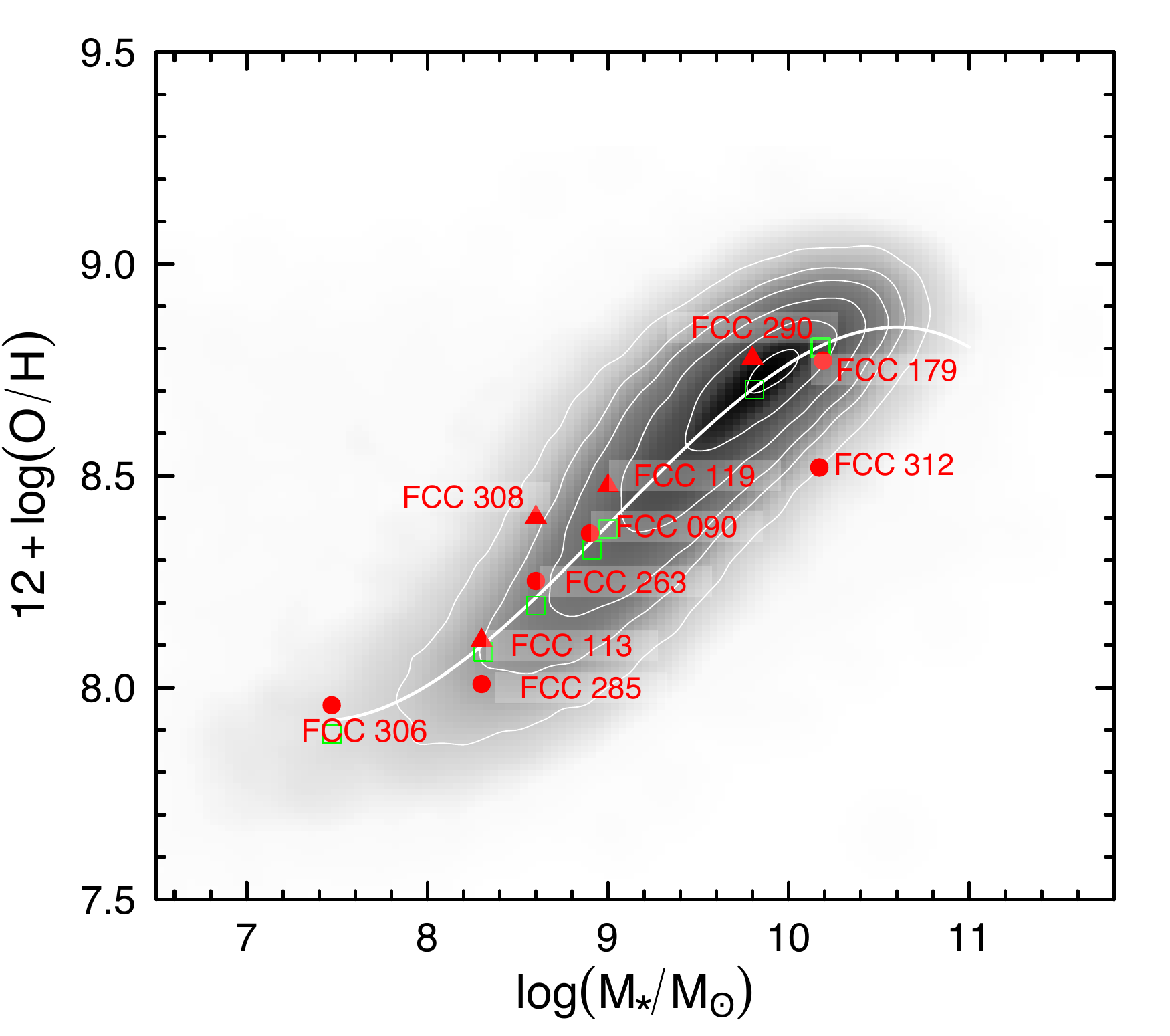}
	\caption{M-Z relation for SDSS (grey background) and Fornax cluster (red symbols) galaxies.  The white solid line indicates the best fit to the SDSS data. Circles and triangles correspond to recent and intermediate infallers, respectively.  The green squares correspond to the control sample.}
        \label{fig:MZFornax}
\end{figure}

\begin{figure*}[ht!]
	\includegraphics[width=0.5\textwidth]{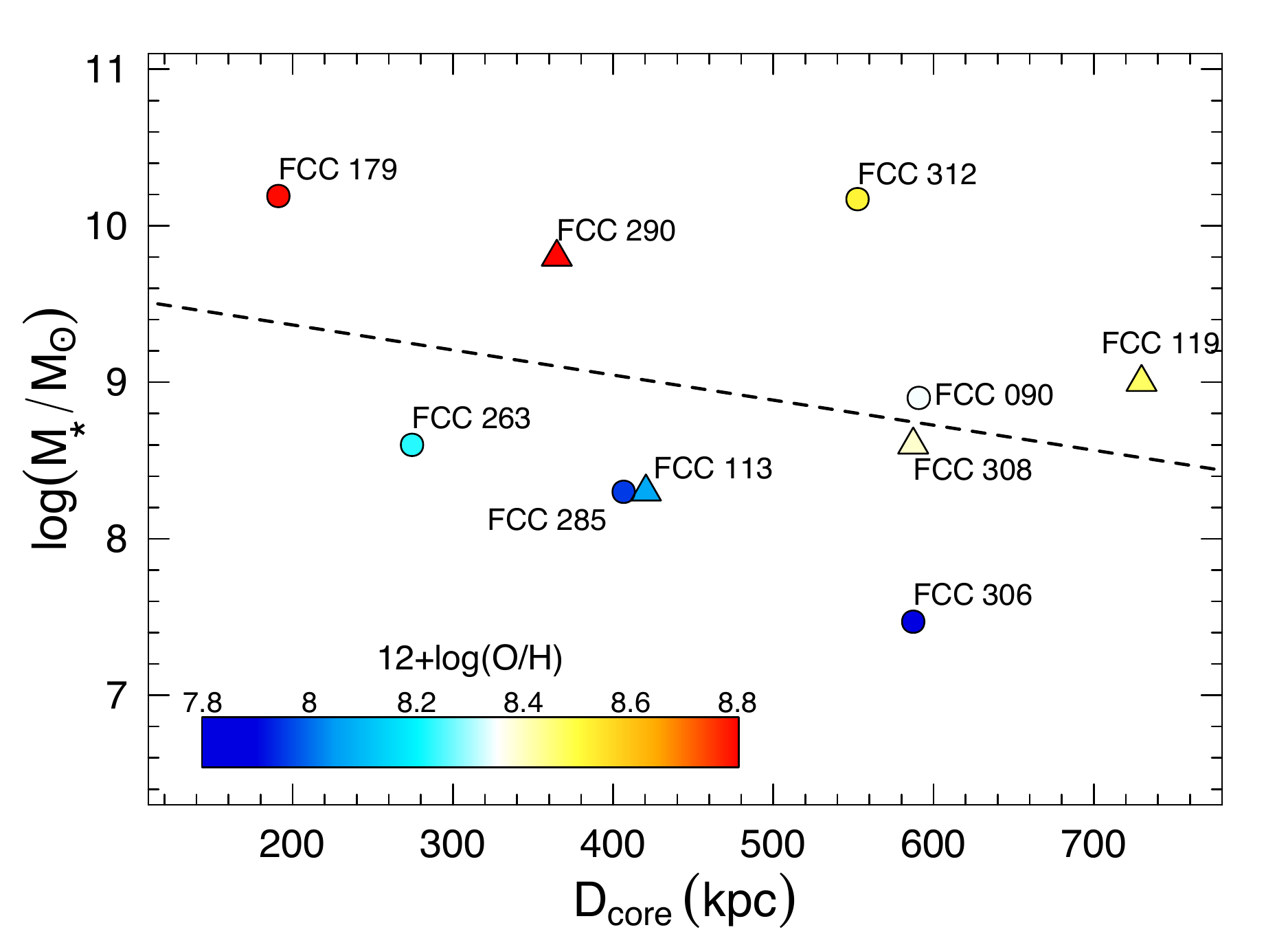}
	\includegraphics[width=0.5\textwidth]{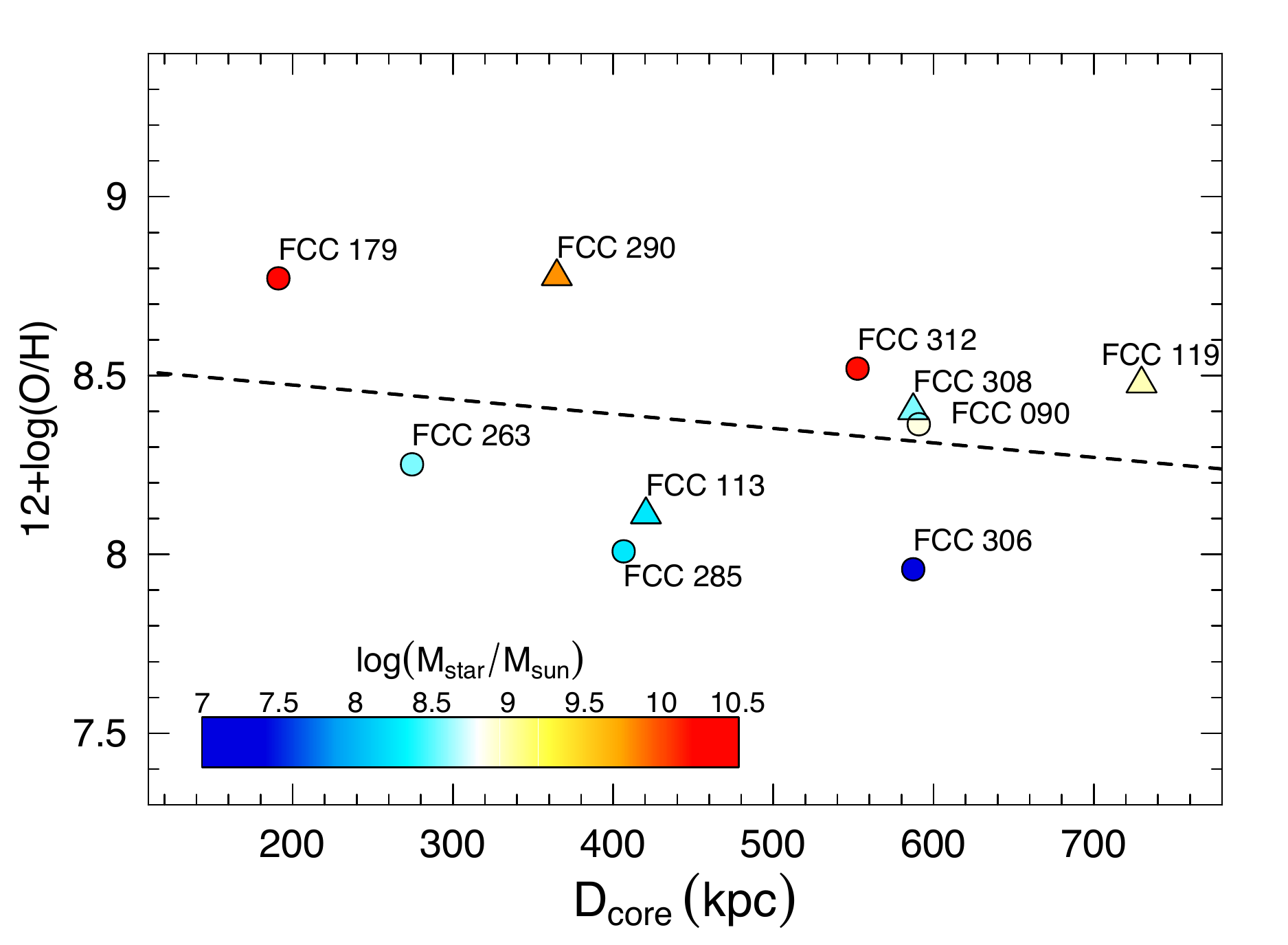}
	\caption{Projected distance-stellar mass relation (left panel), and distance-gas metallicity relation (right panel) for Fornax ELGs. Circles and triangles correspond to recent and intermediate infallers, respectively}
        \label{fig:MDistFornax}
\end{figure*}

To locate our ELGs sample in the context of the global M-Z relation, we integrated the emission lines fluxes from the SF spaxels, and estimated the integrated gas metallicity through Eq. \ref{Eq.Dopita}. The integrated gas metallicities of our sample are listed in Table \ref{tab:Properties}. Next, we constructed the M-Z relation with the integrated values  and total stellar masses, as shown in Fig. \ref{fig:MZFornax}. 

As a comparison, in the same figure we show the M-Z relation for a sample of  $\sim$91,400 galaxies from the  SDSS-DR7 \citep{Abazajian09}. Emission line fluxes of the SDSS galaxies were taken from the OSSY catalog\footnote{https://data.kasi.re.kr/vo/OSSY/index.html} \citep{Oh11}. We selected only SF galaxies using the BPT diagram and the gas metallicities were estimated using the \citet{Dopita16} calibration.  According to \citet{Brinchmann04}, flux measurements become non-Gaussian below a signal-to-noise ratio (SNR) $\sim$ 2, hence we imposed a SNR $>$ 3 for all  the emission lines used. 
Additionally, the SDSS spectra suffer from an aperture bias due to the 3-arcsec diameter fiber used. Since the integrated metallicities tend to be smaller than nuclear ones, we could overestimate the gas metallicities. To minimise the aperture bias, we selected only galaxies that have at least 20$\%$ of their total luminosity inside the SDSS fiber, which should approximate to the integrated value according to \citet{Kewley05}.

We fit (using robust regression) the M-Z relation to the SDSS-SF sample and find the expression: 12$+$log(O/H)$=$ a x$^3$ $+$ b x$^2$ $+ $ c x $+$ d, with the coefficients
a $=$ $-$0.0573 ($\pm$0.0007); b $=$ 1.5478 ($\pm$0.0197); c$=$ $-$13.5028 ($\pm$0.1786); and d$=$ 46.2986 ($\pm$0.5375).

 In addition, we create a control sample by removing all SDSS galaxies in groups from the general SDSS-SF sample described above. We use the catalog of \citet{Tempel12} for the SDSS-DR8, who used the friends-of-friends (FoF) method and identified groups based on their richness. For our purposes, we exclude from our SDSS-SF sample all galaxies with a richness greater than or equal to 2, to ensure we are excluding galaxy pairs as well as groups and clusters. Our final control sample is formed by 60,947 field galaxies. From this sample, we select galaxies within $\pm$0.05 dex in log(M$_\star$/M$_{\sun}$) of each Fornax galaxy, and estimated the median mass and metallicity, shown in green squares in Fig. \ref{fig:MZFornax}. Finally, we estimate the difference (Fornax-control) in metallicity, and obtained a median difference of 0.045 dex for the whole sample.

Galaxies from the Fornax cluster follow the general M-Z relation for SDSS galaxies, and show a statististical 0.045 dex offset towards higher metallicities (see Fig. \ref{fig:MZFornax}). The exception is FCC 312, which shows a low gas metallicity for its stellar mass. 
Previous works report slightly higher ($\sim$ 0.05 dex) gas metallicities for SF galaxies in clusters \citep[e.g.,][]{Ellison09, Gupta16}, explained by the accretion of pre-enriched gas \citep[e.g., ][]{Gupta18}.  Even though our sample is small, our difference of 0.045 dex is consistent with these studies.

\begin{table*}
	\centering
	\caption{Properties of the Fornax ELGs. From left to right, galaxy name, total stellar mass, integrated gas metallicity, projected distance to core center in units of degrees and in kpc, and effective radius in the $r$ band in units of arcsec and kpc. }
	\label{tab:Properties}
	\begin{tabular}{lcccccc} % four columns, alignment for each
		\hline
		Galaxy & log(M$_\star$/M$_{\sun}$) & 12+log(O/H) &  D$_{\rm core}$ (deg) &  D$_{\rm core}$ (kpc) &  R$_{\rm e}$ (arcsec) & R$_{\rm e}$ (kpc)\\
		\hline
		FCC 179 &  10.19 & 8.77  &  0.55 & 191.13&  30.03  & 2.89\\
		FCC 312  & 10.17  & 8.52  &  1.59  & 552.53& 109.5  &  10.57 \\
		FCC 290  & 9.8  & 8.77  &  1.05 & 364.88 &  48.52  & 4.68 \\
		FCC 119 & 9.0  & 8.47  &  2.10 & 729.76 &  17.4  & 1.68 \\
		FCC 090 & 8.9  & 8.36  &  1.70 & 590.76 &  12.1  &  1.167\\
		FCC 263 &  8.6 &  8.25 &  0.79  & 274.53 & 27.15 & 2.62 \\
		FCC 308  & 8.6  &  8.40&  1.69 & 587.28&  37.11   & 3.58 \\
		FCC 113 & 8.3 &  8.11 &  1.21 & 420.48 & 20.56 &  1.98 \\
		FCC 285 &  8.3 & 8.0  &  1.17   & 406.58 & 49.9  & 4.82\\
		FCC 306  & 7.47  & 7.95  &  1.69 & 587.28 &  9.7  & 0.94 \\
		\hline
	\end{tabular}
\end{table*}

%\begin{table}
%	\centering
%	\caption{This is an example table. Captions appear above each table.
%	Remember to define the quantities, symbols and units used.}
%	\label{tab:example_table}
%	\begin{tabular}{lccc c} % four columns, alignment for each
%		\hline
%		Galaxy & Stellar mass & Re$_r$ (arcsec) & PA$_{\rm kin}$ & D$_{\rm core}$ (deg)\\
%		\hline
%		FCC113 & 8.3 & 20.56 & 2 & 1.21\\
%		FCC179 &  10.19 & 30.03 & & 0.55\\
%		FCC263 &  8.6 & 27.15 & & 0.79\\
%		FCC285 &  8.3 & 49.9 & & 1.17 \\
%		FCC290  & 9.8  & 48.52 &. & 1.05 \\
%		FCC306  & 7.47  & 9.7 & & 1.69 \\
%		FCC308  & 8.6  & 37.11 & & 1.69 \\
%		FCC312  & 10.17  & 109.5 & & 1.59 \\
%		FCC090* & 8.9  &  12.1& 72 & 1.7 \\
%		FCC119* & 9.0  & 17.4  & 43 &  \\
%		FCC184* & 10.67  &  35.5 & 54 &  0.31\\
%		
%		\hline
%	\end{tabular}
%\end{table}
%

%##Metallicity as a distance to Core

\subsection{Stellar mass and metallicity segregation}\label{subsec:MassSegregation}

Another distinctive feature in galaxy clusters is  stellar mass segregation \citep{Chandrasekhar43}, or the tendency of more massive galaxies to be located closer to the center \citep[e.g., ][]{DeLucia04,Contini12,Kim20}. Nonetheless, it is still controversial, with some authors finding observational and theoretical evidence for no segregation \citep[e.g., ][]{Lares04,Balogh14,Roberts15,Nascimento17}, or weak segregation \citep[e.g.,][]{vonderLinden10, Ziparo13,Vulcani13, Joshi17}. 

Recently, \citet{Kim20} concluded that mass segregation is more visible in low-mass clusters simply as a result of the shorter dynamical friction time  for more massive galaxies. Since the Fornax cluster can be considered a low mass cluster \citep[M$_{\rm vir}$ $\sim$ 7 $\times$ 10$^{13}$ M$_{\odot}$,][]{Drinkwater01}, we investigate whether or not mass segregation is present.  Figure \ref{fig:MDistFornax} shows the projected distance vs. stellar mass relation for our Fornax-ELGs cluster sample. Although with a large scatter,  with a Pearson correlation coefficient of -0.3 and a corresponding significance of $p=0.4$ of its deviation from the null hypothesis, our data suggest a mild relation as described by Eq. \ref{Eq:MassDist}.

\begin{equation}\label{Eq:MassDist}
%{\rm log(M_\star/M_{\sun})} =  0.001602 (\pm 0.001788) \times D_{\rm core} + 9.6866 (\pm 0.8874)
{\rm log(M_\star/M_{\sun})} =  0.0016 (\pm 0.0017) \times D_{\rm core} + 9.686 (\pm 0.887) 
\end{equation}

A mass segregation in clusters would naturally produce a similar effect in the galaxy metallicities due to the M-Z relation, and the gas metallicity would decrease as a function of projected clustercentric distance. We explore this in Fig. \ref{fig:MDistFornax}, and find a relation described by Eq.  \ref{Eq:MetDist}.  Again, our data shows a slight correlation, with a Pearson coefficient of -0.23 and a value $ p=0.5$.

\begin{equation}\label{Eq:MetDist}
{\rm 12+log(O/H)} =  -0.0004 (\pm 0.0005) \times D_{\rm core} + 8.554 (\pm 0.263) 
\end{equation}

Moreover, our sample shows a mild signature of both, mass and metallicity segregation.  Furthermore, by using the early type population of F3D galaxies, \citet{Spriggs21} find signs of mass segregation.

\section{Results}\label{sec:Results}

\subsection{Gas metallicity gradients}\label{subsec:MetGradients}

\begin{figure}
\includegraphics[width=0.45\textwidth]{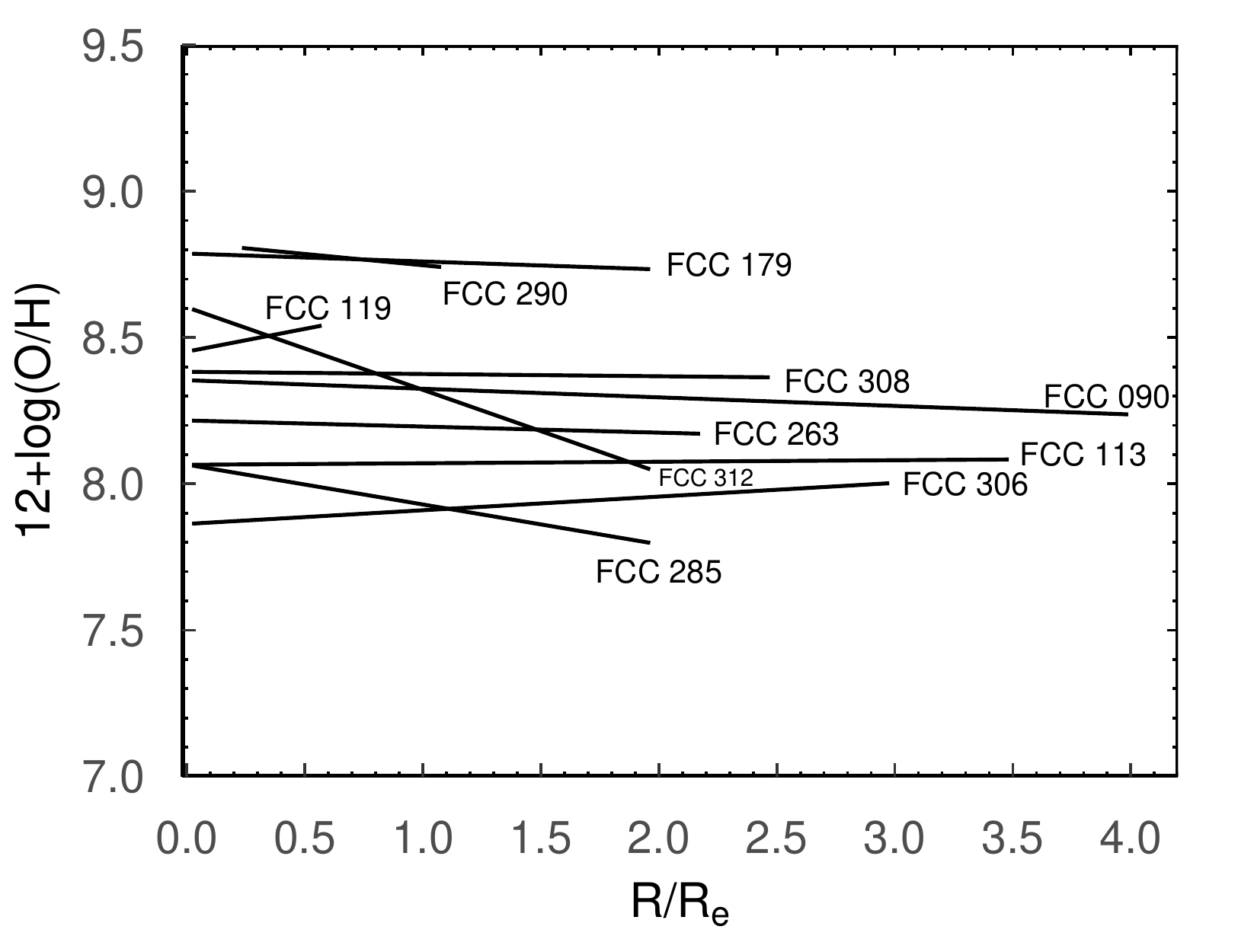}
    \caption{Linear fits of the gas metallicity radial profiles for Fornax-ELGs sample.}
    \label{fig:MetGradientsAll}
\end{figure}

In this section we explore the effect of cluster environment on the gas metallicity gradients of galaxies. We used the resolved gas metallicity estimations for every MUSE spaxel, described in Sect \ref{sec:SampleCharacterization}. Metallicity gradients are presented in the literature either as a function of kpc scales, normalized as a function of  R25, or  R$_{\rm e}$. To compare our results with previous works in the literature and data from IllustrisTNG, we opted to use  R$_{\rm e}$.  All radial distances in the gradients are corrected for inclination and normalised to R$_{\rm e}$ in r-band \citep{Raj19}, given in Table \ref{tab:Properties}. The resulting metallicity gradients for all our Fornax ELGs sample are shown in Fig. \ref{fig:MetGradientsAll}.  All the linear fits are performed, over the full radial extent of each galaxy using all the spaxels, in the language and environment for statistical computing ``R'' with the package ``HYPERFIT'' \citep{Robotham2015}.  We define the root-mean-square error (RMSE) as $\sqrt{\sum_{i=1}^{n} (\hat{y}_i-y_i)^2 / n }$, where $\hat{y}_i$  and $y_i$ are the predicted and observed values, respectively. The  slope, intercept and RMSE of the metallicity gradients  are given in Table \ref{tab:SummaryFCC}.

To identify any systematic change due to the cluster environment, we proceed to create a control sample as detailed in the next section.

\subsection{Control sample}\label{subsec:control}

Previous papers looking for environmental effects  used the stellar mass of galaxies as a baseline to create control samples, together with other properties such as color or morphology \citep[e.g., ][]{Ellison08,Ellison09,Scudder12,Garduno21,SotilloRamos21}. 
Due to the limited number of ELGs with IFU data, we are focussing on the stellar mass as a reference, since it is one of the most fundamental galaxy properties, and scales with both, star formation rate (SFR) \citep[e.g., ][]{Brinchmann04,Noeske07,Elbaz07} and metallicity \citep[e.g., ][]{Tremonti2004}.

From the plethora of current IFU surveys, we use galaxies from the MAD survey \citep[][]{Erroz19}, also observed with MUSE, and hence offer a direct comparison with our data. We select the SF galaxies from their public database\footnote{https://www.mad.astro.ethz.ch/} that are within $\pm$0.2 dex  the stellar mass of each galaxy in our  Fornax sample.  The current public MAD data provides galaxies only with stellar masses higher than $\sim$10$^{8.4}$ M$_\star$/M$_{\sun}$.
Our final comparison sample from MAD  is formed by 16 galaxies. Next, we estimate metallicities using the same prescription used for Fornax ELGs, and normalized the metallicity gradients to their respective R$_{\rm e}$ for a direct comparison. Each linear fit on the galaxies of the control sample was performed as well with HYPERFIT. 

\begin{figure*}
\includegraphics[width=0.33\textwidth]{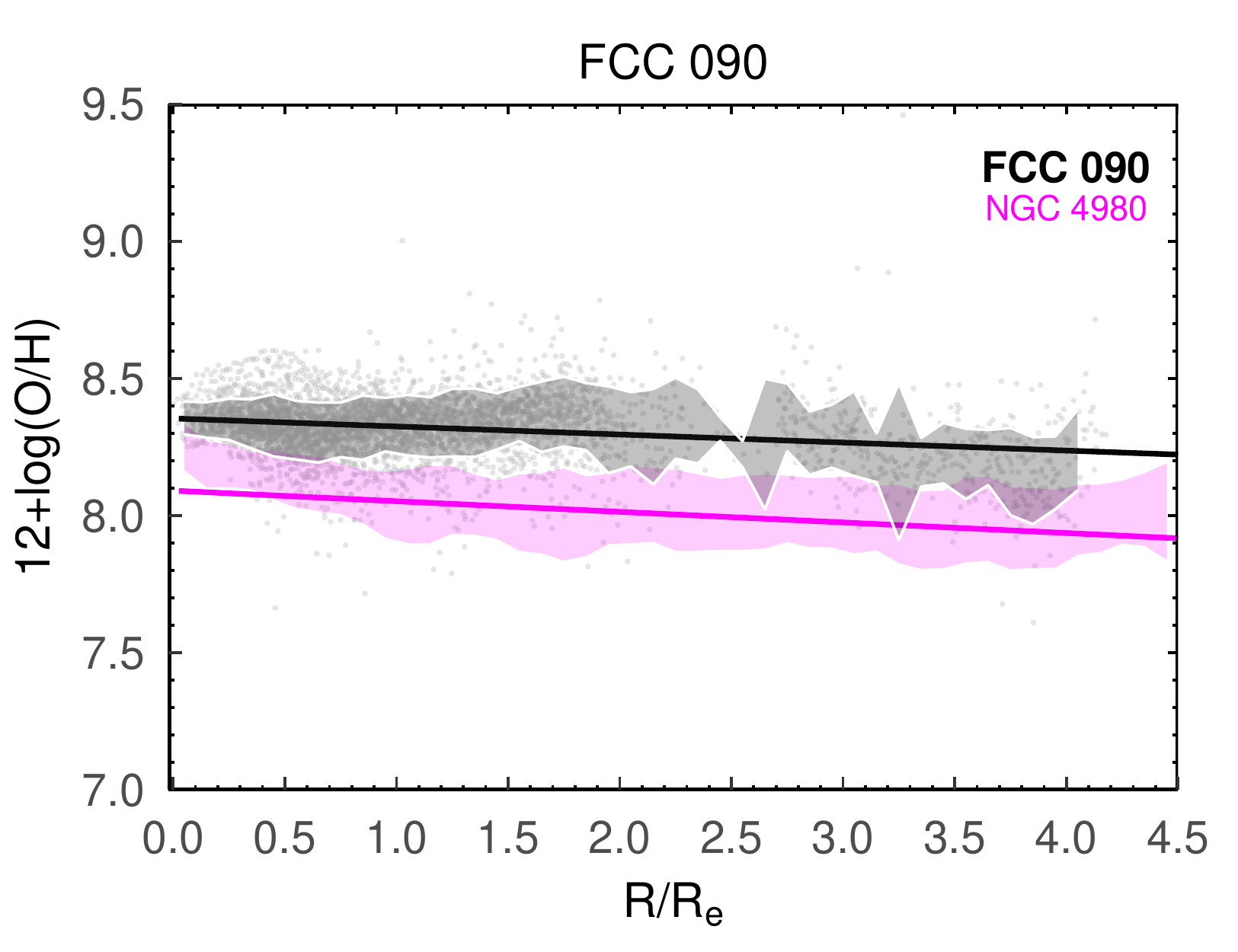}
\includegraphics[width=0.33\textwidth]{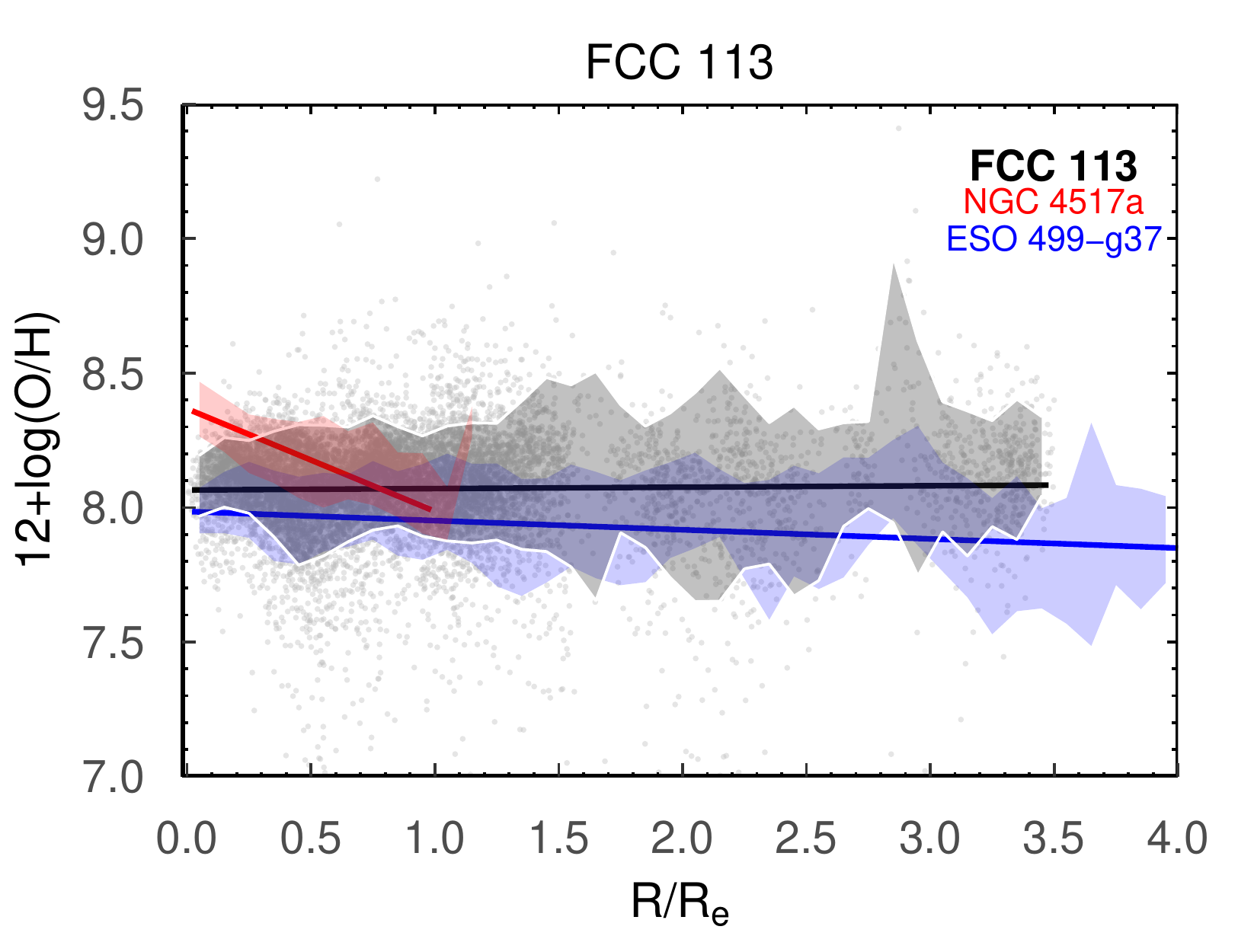}
\includegraphics[width=0.33\textwidth]{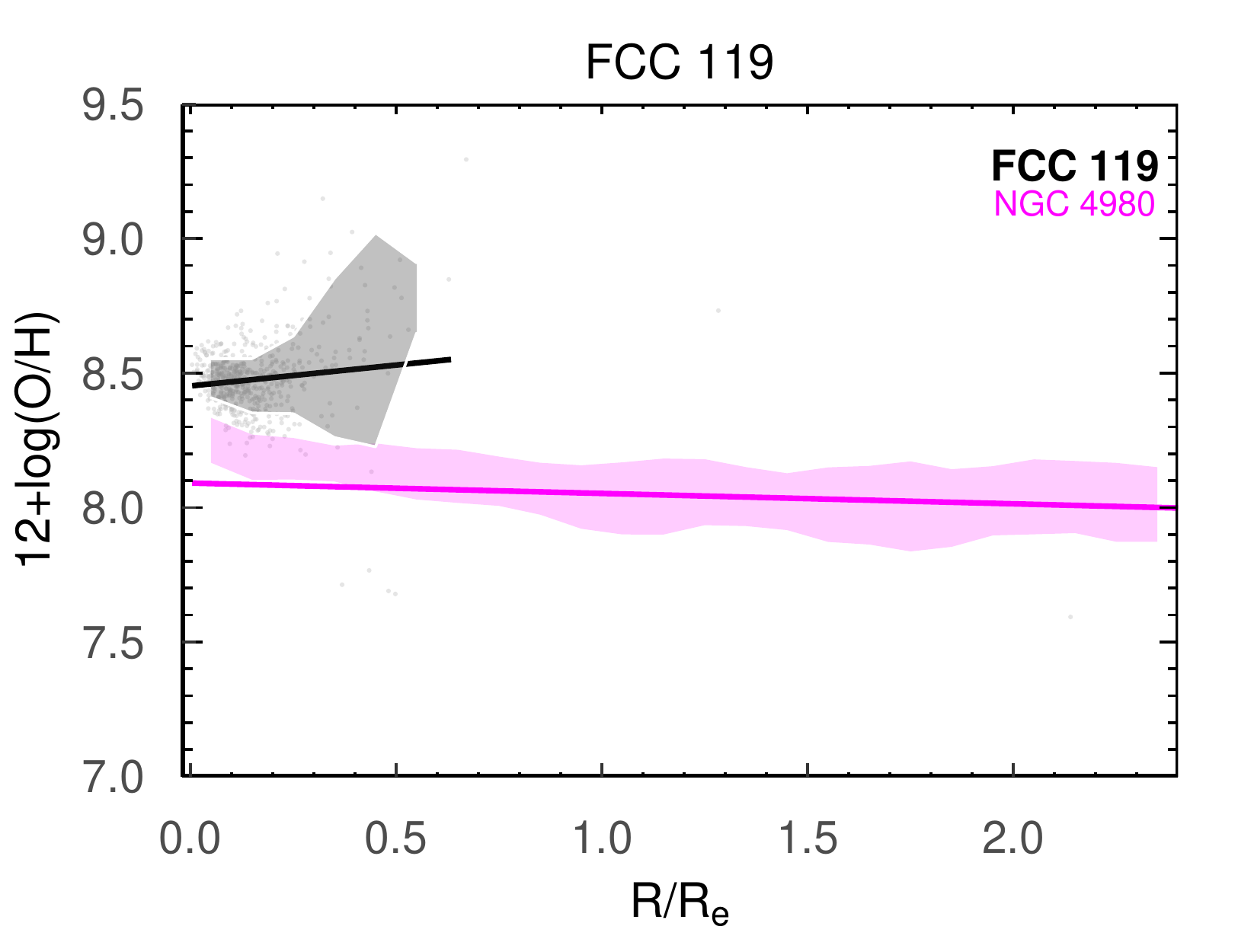}
\includegraphics[width=0.33\textwidth]{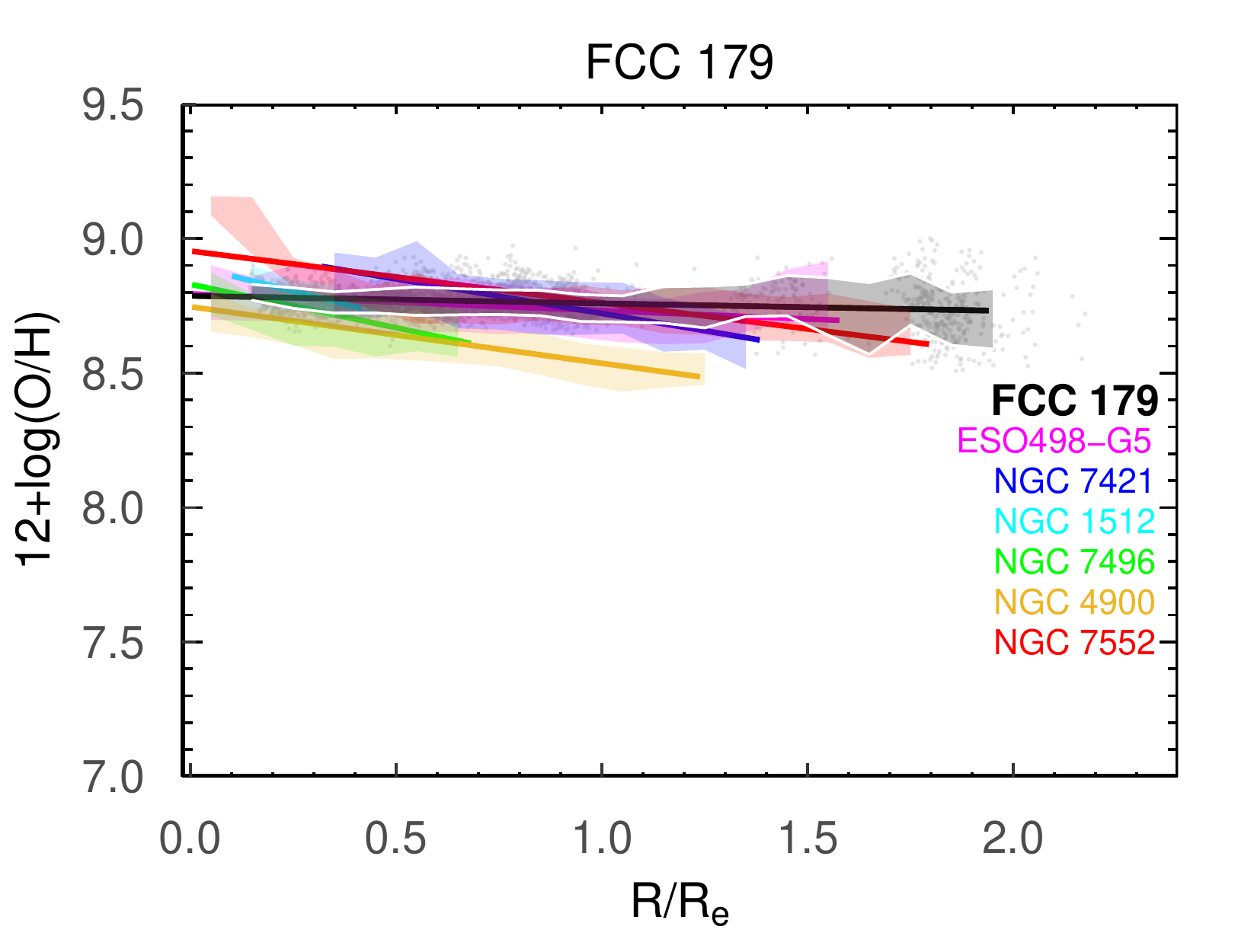}
\includegraphics[width=0.33\textwidth]{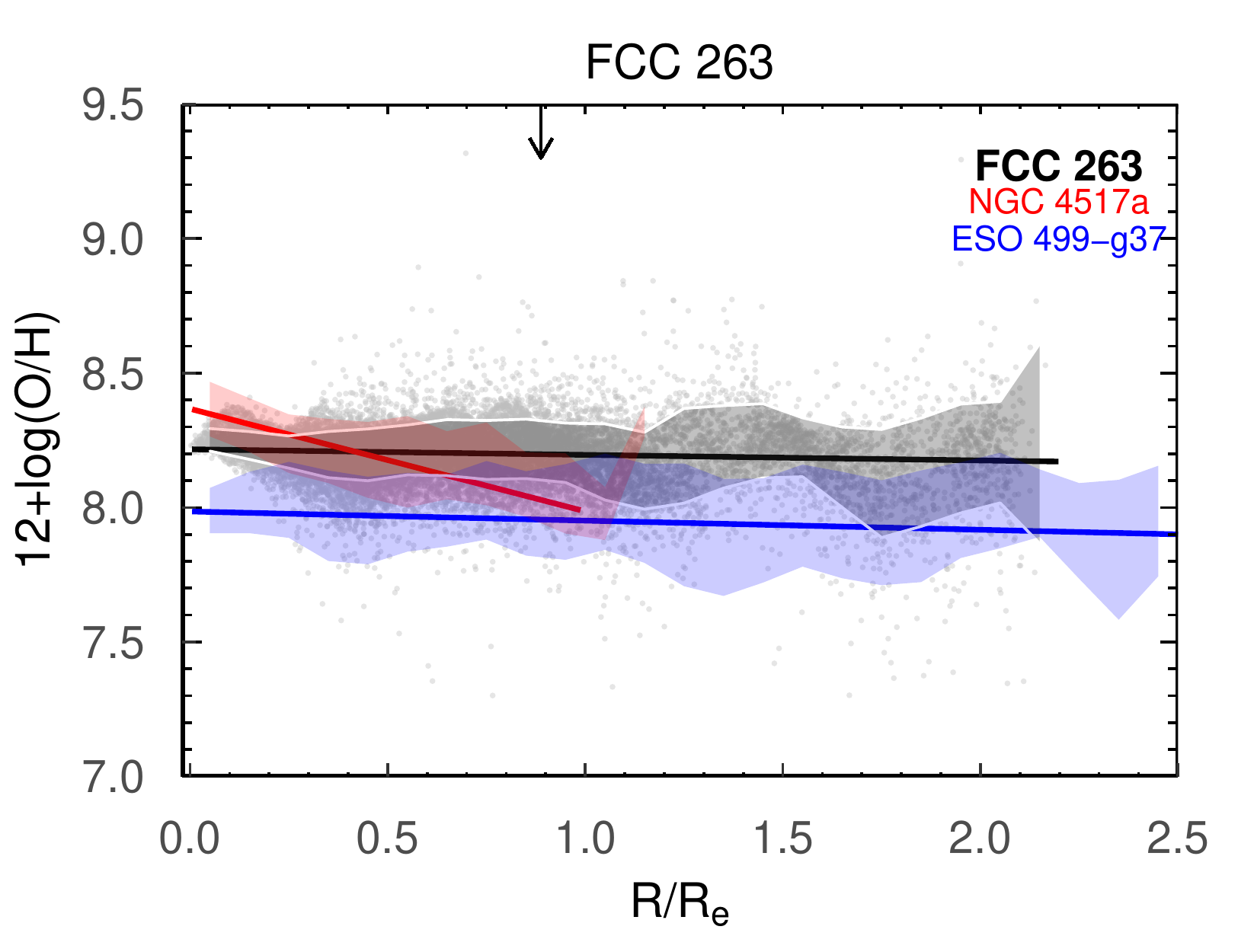}
\includegraphics[width=0.33\textwidth]{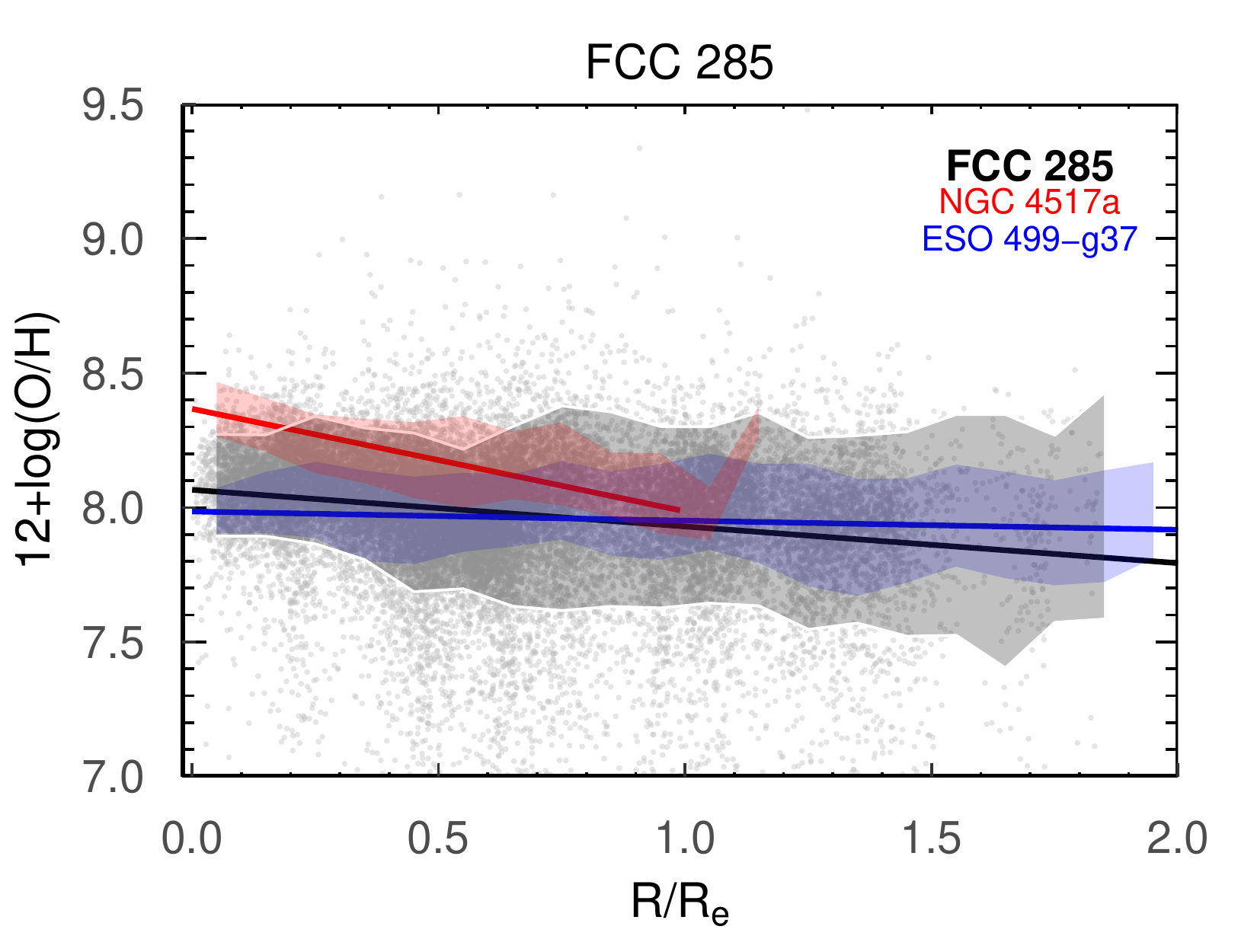}
\includegraphics[width=0.33\textwidth]{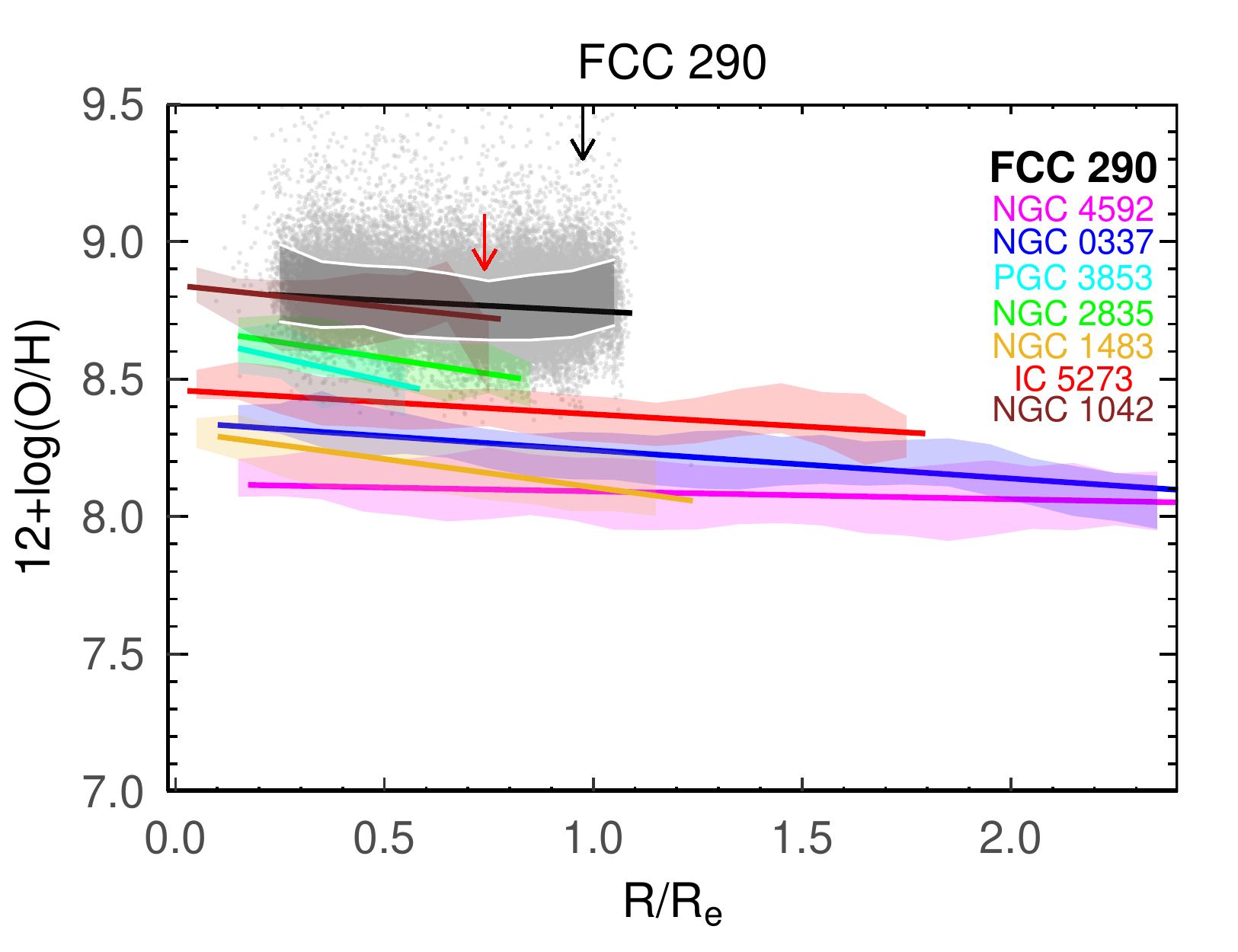}
\includegraphics[width=0.33\textwidth]{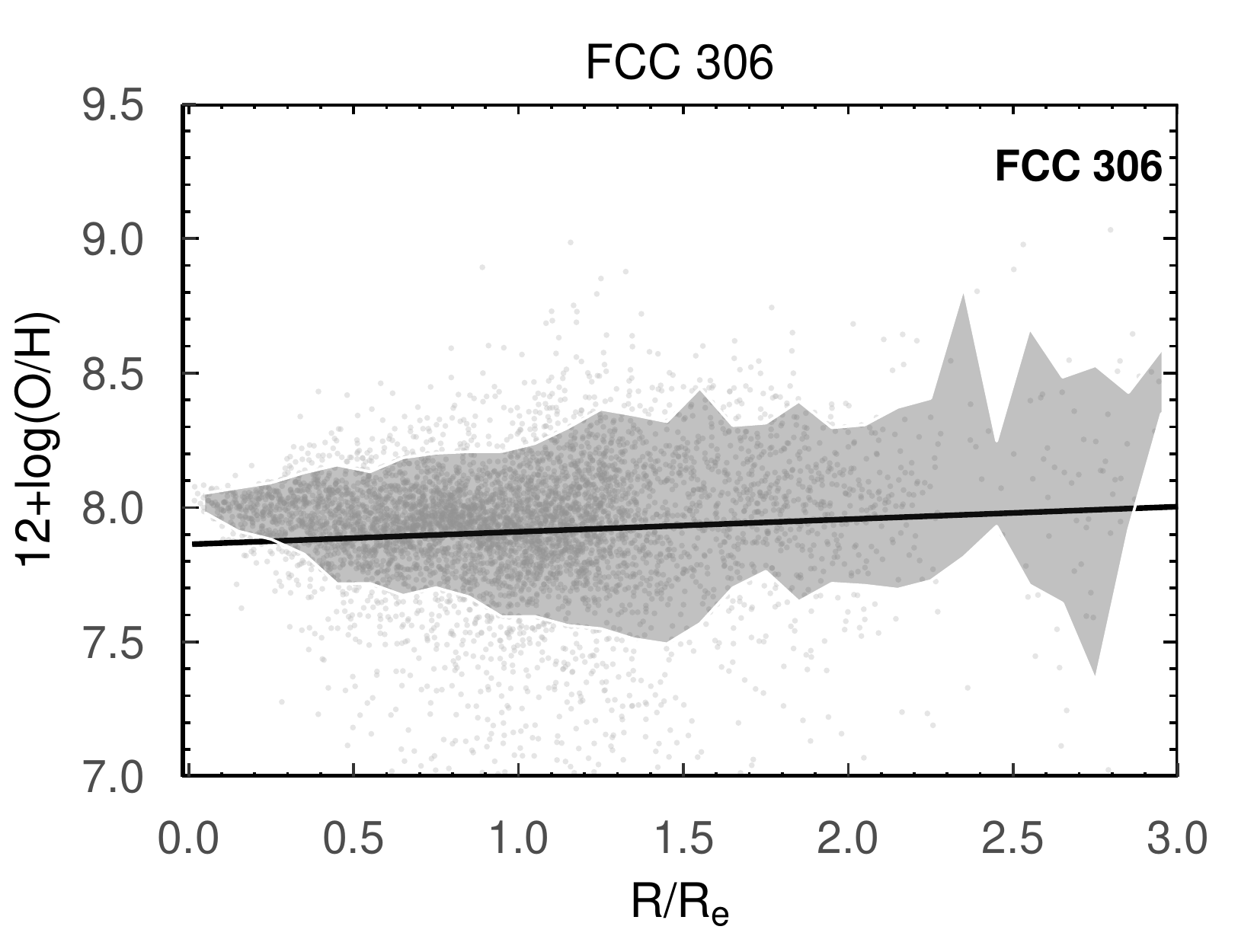}
\includegraphics[width=0.33\textwidth]{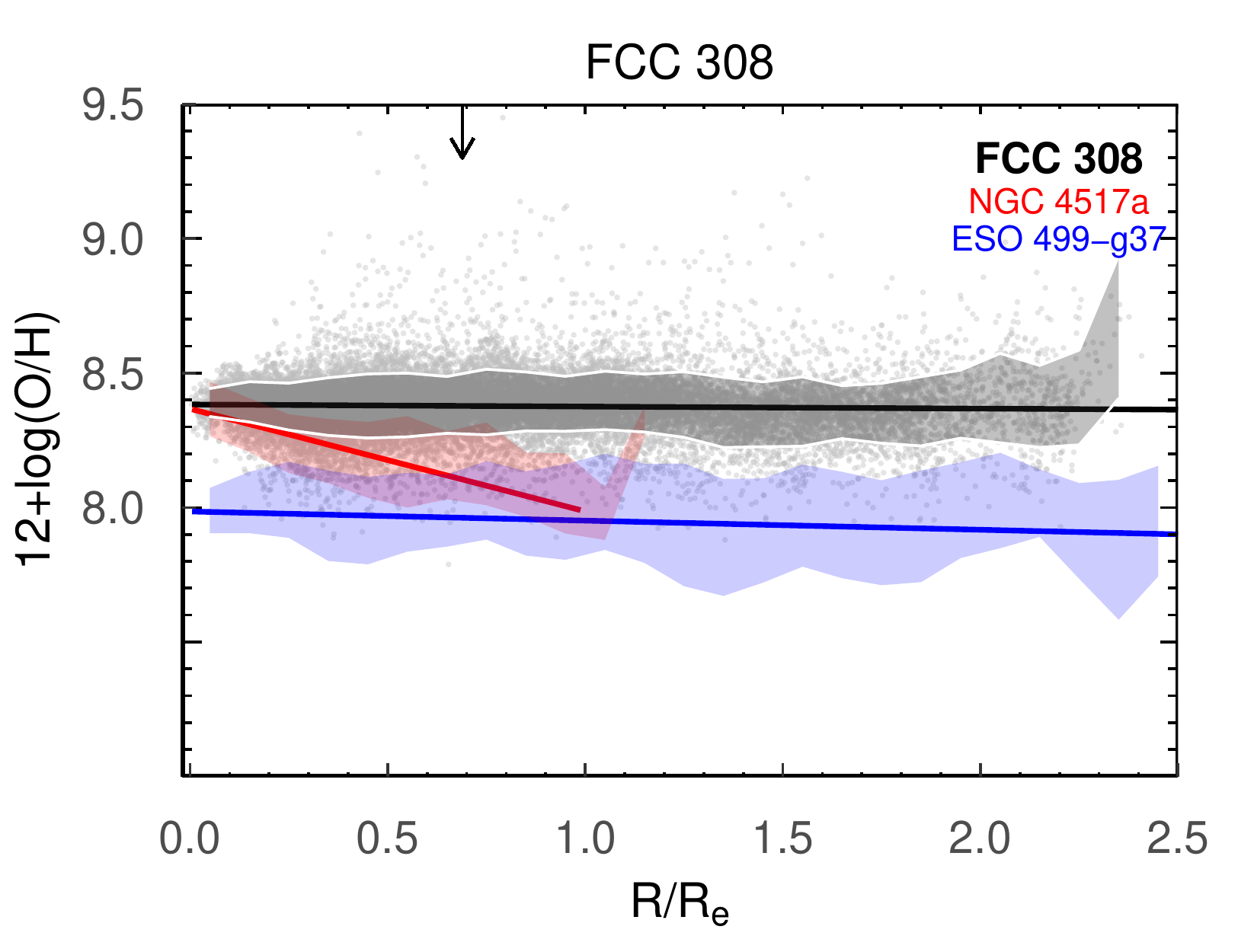}
\includegraphics[width=0.33\textwidth]{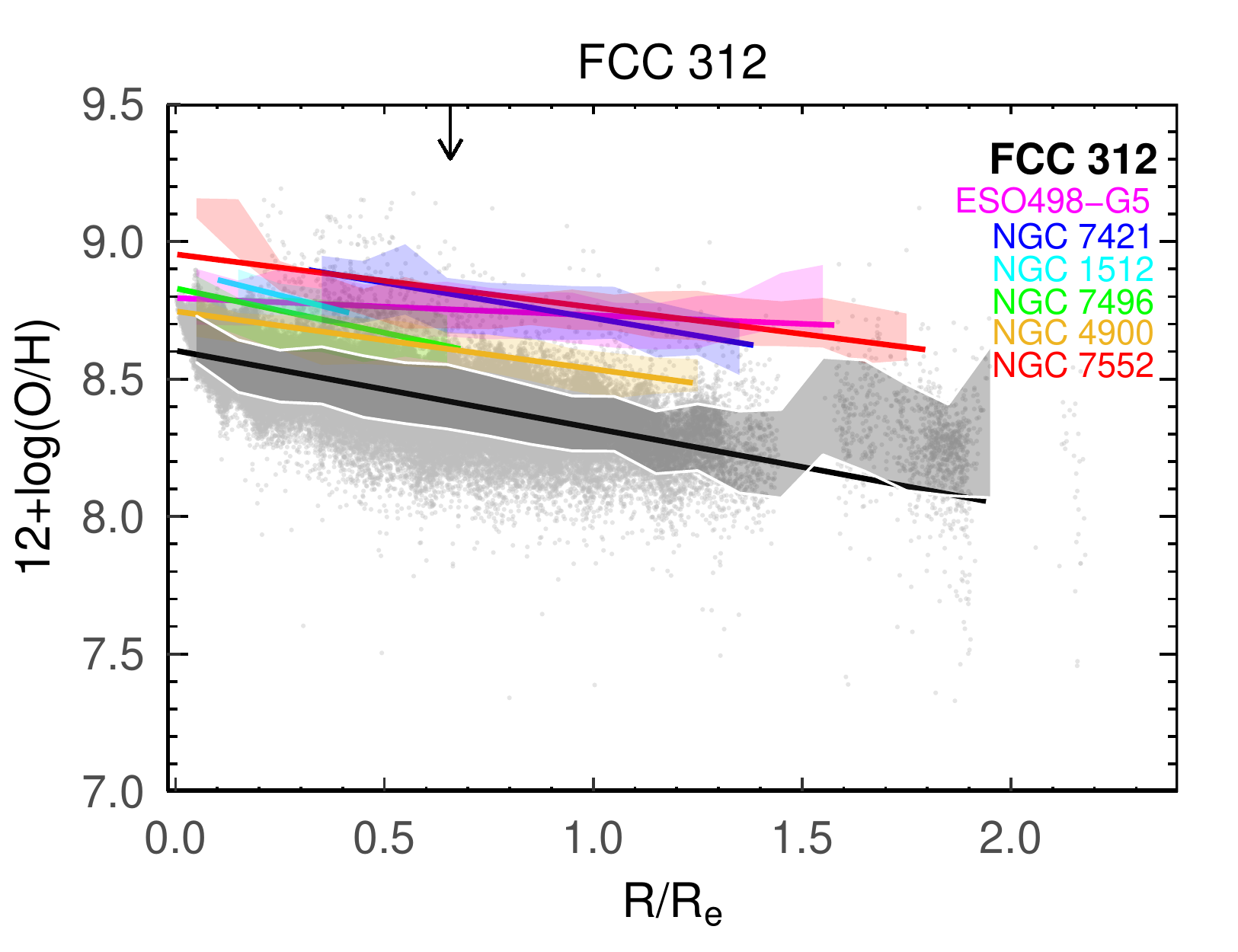}
%This is to get low Resolution size 
%\includegraphics[width=0.33\textwidth]{FCC090.pdf}
%\includegraphics[width=0.33\textwidth]{FCC113.pdf}
%\includegraphics[width=0.33\textwidth]{FCC119.pdf}
%\includegraphics[width=0.33\textwidth]{FCC179.pdf}
%\includegraphics[width=0.33\textwidth]{FCC263_Arrow.png}
%\includegraphics[width=0.33\textwidth]{FCC285.png}
%\includegraphics[width=0.33\textwidth]{FCC290_Arrow.pdf}
%\includegraphics[width=0.33\textwidth]{FCC306.pdf}
%\includegraphics[width=0.33\textwidth]{FCC308_arrow.png}
%\includegraphics[width=0.33\textwidth]{FCC312_Arrow.png}
    \caption{Gas metallicity radial profiles for our sample of Fornax ELGs. Each panel corresponds to a Fornax galaxy (grey dots) and its corresponding control sample from the MAD survey (colour dots). The shaded areas around each gradient correspond to the 1 $\sigma$ dispersion in 0.2 dex of R/R$_{\rm e}$.   
    The gradients of the control samples are coloured according to the names inside each panel. The black vertical arrow in the panels of FCC 263, FCC 290, FCC 308, and FCC 312 shows the break radius from \citet{Raj19}.  The red arrow in the panel of FCC 290 shows the drop in H$_2$-to-dust ratio from \citet{Zabel21}.}
    \label{fig:PanelsMetGradientsWithControl}
\end{figure*}

The metallicity vs. galactocentric distance relation for individual Fornax ELGs and their corresponding control sample from MAD is shown in Fig. \ref{fig:PanelsMetGradientsWithControl}. In each panel the shaded regions show the dispersion of $\pm$1 $\sigma$ in bins of 0.2 R/R$_{\rm e}$.  
%{\bf In addition, for four of our galaxies we indicate with a vertical arrow the "break radius", derived from surface brightness profiles from \citet{Raj19}. The rest of our sample show either no break radius, or it lies outside the area observed by MUSE. The break radius indicates the radius where the logarithmic surface brightness of a galaxy shows a discontinuity. Our data shows no relation between the metallicity gradients and the break radius, suggesting that it might be more related to a break in star formation as suggested by \citet{Raj19}.}
The metallicity gradients for Fornax and MAD galaxies are shown in Fig. \ref{fig:MetGradientsWithControl}, where the vertical lines indicate the RMSE of the fit for both samples.

We also take into account the metallicity gradients from the SAMI survey \citep{Croom12}, since SAMI observed low mass galaxies, which are difficult to find in other IFU surveys such as MANGA, CALIFA, and MAD. For SAMI galaxies, the gradients were taken from \citet{Poetrodjojo21}. From this paper, we selected the gradients estimated using the metallicity method of  \citet{Dopita16}, and also normalized to the R$_{\rm e}$. All the metallicity gradients from the SAMI sample   are shown in Fig. \ref{fig:MetGradientsWithControl} to illustrate the general trend with stellar mass.
%The SAMI sample is the only one that provide galaxies with a similar  mass  range as FCC 306 (log(M$_\star$/M$_{\sun}$) = 7.47).

To increase our control sample in the low mass regime, we use a couple of galaxies from  \citet[][hereafter B19]{Bresolin19}.  The gas metallicities were estimated following \citet{Dopita16}, and the corresponding gradients were normalized to R$_{\rm e}$. The obtained gradients are shown coloured in cyan in Figure \ref{fig:MetGradientsWithControl}. 

In addition, we look for any effect of the break radius on the metallicity gradient for Fornax late-type galaxies.  \citet{Raj19} estimated the break radius as the radius where the logarithmic surface brightness of a galaxy shows a discontinuity. We indicate the break radius for four of our galaxies as a vertical arrow in Fig. \ref{fig:PanelsMetGradientsWithControl}.  The rest of our sample show either no break radius, or it lies outside the area mapped by MUSE. Our data shows no relation between the break radius and the metallicity gradient, suggesting that it might be more related to a break in star formation as suggested by \citet{Raj19}.  In  Fig. \ref{fig:PanelsMetGradientsWithControl}, the red arrow in the FCC 290 panel shows the drop in H$_2$-to-dust ratio from \citet{Zabel21}. Interestingly, this drop suggests a change of the metallicity gradient towards a flatter/positive gradient.

\subsection{Cluster-Control comparison}\label{subsec:comparison}

\begin{figure*}
\includegraphics[width=0.83\textwidth]{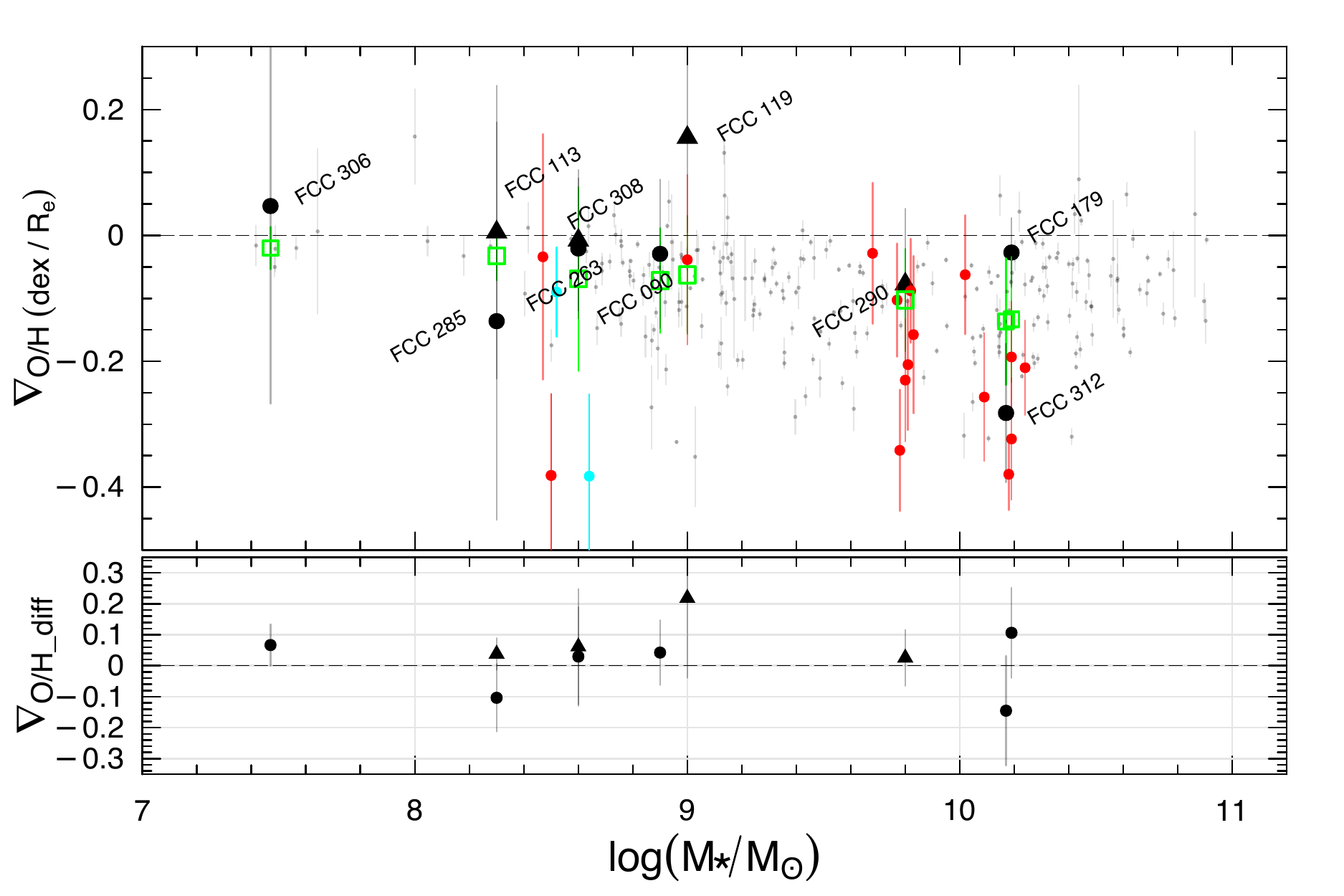}
    \caption{ Gas metallicity gradients as a function of stellar mass. The black symbols show our Fornax ELGs sample, with circles and triangles corresponding to recent and intermediate infallers, respectively. The red symbols correspond to galaxies from the MAD survey. The small grey circles indicate galaxies from the SAMI survey. The cyan circles are galaxies from B19. For galaxies from Fornax, MAD, and B19, the vertical bars correspond to the RMSE. For SAMI galaxies, the bars indicate the 1 $\sigma$ error in the metallicity gradient. The green squares show the median values of $\tilde{\mathcal \nabla}$$_{\rm O/H}$ and their 1 $\sigma$ dispersion for the control sample (MAD, SAMI and Bresolin et al.)  around the stellar mass of each Fornax galaxy, see text for details.  The bottom panel shows the difference in metallicity gradient between each Fornax galaxy and the corresponding  value $\tilde{\mathcal \nabla}$$_{\rm O/H}$ from the control sample.}
    \label{fig:MetGradientsWithControl}
\end{figure*}

\begin{figure*}
\includegraphics[width=0.83\textwidth]{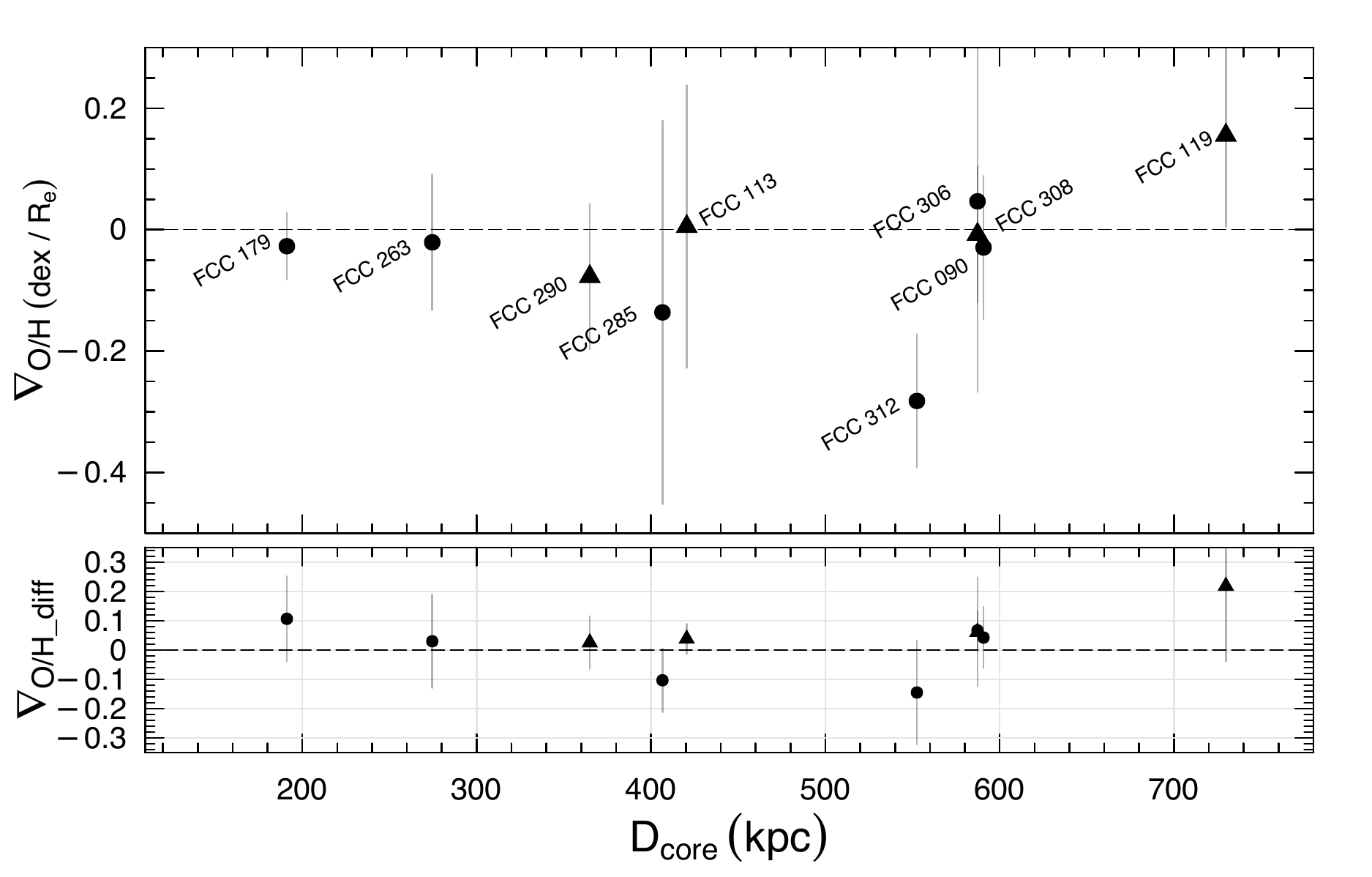}
    \caption{Gas metallicity gradients as a function of the projected cluster-centric distance to the BCG. The bottom panel shows the difference of gas metallicity gradient between each Fornax galaxy and their corresponding $\tilde{\mathcal \nabla}$$_{\rm O/H}$  value from the control sample.}
    \label{fig:MetGradientsWithDistance}
\end{figure*}

Figure \ref{fig:MetGradientsWithControl} shows the direct comparison between the Fornax metallicity gradients and control galaxies from MAD, SAMI, and  B19. As described above, all the gradients  were measured using the same metallicity method, and normalized to R$_{\rm e}$. The vertical bars show the RMSE of the fit for Fornax, MAD and B19 galaxies, while for the SAMI galaxies they correspond to  the 1 $\sigma$ error in the gradient. 

To identify systematic changes in the metallicity gradients of Fornax ELGs, we combine all the control galaxies (from MAD, SAMI, and B19) and estimate the median value ($\tilde{\mathcal \nabla}$$_{\rm O/H}$) for all galaxies within $\sim$$\pm$0.2 dex  in log(M$_\star$/M$_{\sun}$) of each Fornax galaxy. The values $\tilde{\mathcal \nabla}$$_{\rm O/H}$  are indicated as green squares in Fig. \ref{fig:MetGradientsWithControl}. 

%estimated the medium values ($\tilde{\mathcal \nabla}$$_{\rm O/H}$) by combining all the control galaxies within $\sim$$\pm$0.2 dex  in log(M$_\star$/M$_{\sun}$) of each Fornax galaxy. 

The bottom panel in the same Figure shows the difference between the metallicity gradient of each Fornax ELGs with its corresponding $\tilde{\mathcal \nabla}$$_{\rm O/H}$.  With the exception of FCC 285 and FCC 312, the rest of our Fornax ELGs shows more positive gradients with respect to those of the control sample, either flatter or positive (e.g., FCC 119 and FCC 306).  In fact, FCC 119 also shows a positive stellar metallicity gradient, which is likely caused by the metal-poor nuclear star cluster in the centre of FCC\,119 \citep{Fahrion21}.
 The vertical bars show the 1 $\sigma$ dispersion of the difference. The histogram of these differences (Fig. \ref{fig:GradDiff_F3DAndIllustris}), shows a shift of $\sim$0.04  dex/R$_{\rm e}$ towards flatter gradients for the Fornax sample. 
%To test the probability of obtaining this difference by chance, we performed a bootstrap by constructing samples with replacement 1000 times. From each sample, we created histograms with the differences, as in Fig. \ref{fig:GradDiff_F3DAndIllustris}, and estimated the median of the differences (${\mathcal \nabla}$$_{\rm O/H_{diff}}$). We find that.....  DO THE BOOTSTRAP.
To quantify the probability of obtaining this difference by chance, we bootstrap the data by constructing samples with replacement 1000 times and  find an error of  $\sim$0.02 $\%$ in our difference.
 %The probability of finding this difference by chance is only $\sim$2$\%$.
%We bootstrap the data by constructing samples with replacement 1000 times, and find a probability of $\sim$2$\%$ of finding that difference by chance.

%To explore if the observed flattening could be due to the projected clusterdistance, in Figure \ref{fig:MetGradientsWithDistance}

Considering the possibility that the observed flattening is related to the galaxy distance to the cluster centre, we show in Figure \ref{fig:MetGradientsWithDistance} the metallicity gradients of the Fornax ELGs as a function of the projected distance to the brightest cluster galaxy (BCG) NGC 1399. The lower panel in the same figure shows the difference with the control sample. No pattern emerges, indicating that the projected distance to the BCG does not play a significant role,  although a larger sample is needed to confirm this result, and we do not discard that  accurate 3D distances might change this outcome \citep{Rys14}.
%Further work is necessary to estimate the physical distances to the BCG/center, for instance using distance estimators such as planetary nebulea (THOMAS paper), 

%
%\begin{figure*}
%\includegraphics[width=0.33\textwidth]{FCC312_MetGradientD16.pdf}
%\includegraphics[width=0.33\textwidth]{FCC113_MetGradientD16.pdf}
%\includegraphics[width=0.33\textwidth]{FCC179_MetGradientD16.pdf}
%\includegraphics[width=0.33\textwidth]{FCC263_MetGradientD16.pdf}
%\includegraphics[width=0.33\textwidth]{FCC285_MetGradientD16.pdf}
%\includegraphics[width=0.33\textwidth]{FCC290_MetGradientD16.pdf}
%\includegraphics[width=0.33\textwidth]{FCC308_MetGradientD16.pdf}
%\includegraphics[width=0.33\textwidth]{FCC306_MetGradientD16.pdf}
%\includegraphics[width=0.33\textwidth]{FCC184_MetGradientD16.pdf}
%\includegraphics[width=0.33\textwidth]{FCC119_MetGradientD16.pdf}
%\includegraphics[width=0.33\textwidth]{FCC090_MetGradientD16.pdf}
%    \caption{Metallicity gradient using Dopita et al.}
%    \label{fig:example_figure2}
%\end{figure*}
%

\begin{table*}
	\centering
%	\caption{{\bf List of coefficients and RMSE for the linear fits of the metallicity gradients for Fornax galaxies. The last column indicate if the galaxy is a recent or Intermediate infaller.}}
		\caption{ List of the gas metallicity gradient coefficients and RMSE for our sample of Fornax ELGs. The last column indicates if the galaxy is a recent or intermediate infaller in the Fornax cluster.}

	\label{tab:SummaryFCC}
	\begin{tabular}{lccccccc} % four columns, alignment for each
		\hline
	Galaxy 	&	intercept		&	slope		&	RMSE		&	Infaller\\
		\hline
FCC 119	&	8.452	($\pm$ 0.007)&	0.1555 ($\pm$ 0.0331)	&		0.1506	&		Int\\
FCC 090	&	8.3542	($\pm$ 0.0028)&	-0.0293	($\pm$ 0.0018)&		0.1179	&			Rec\\
FCC 113	&	8.065	($\pm$ 0.003)& 	0.005205	($\pm$ 0.003184)&		0.2328	&		Int\\
FCC 263	&	8.2166	($\pm$ 0.0013)&	-0.0208	($\pm$ 0.0015)&		0.1115	&			Rec\\
FCC 285	&	8.0658	($\pm$ 0.0030)&	-0.1362	($\pm$ 0.0039)&		0.3157	&			Rec	\\
FCC 290	&	8.824	($\pm$ 0.001)&	       -0.07732	($\pm$ 0.00264)&		0.1198	&		Int\\
FCC 306	&	7.863	($\pm$ 0.011)&	     0.04669 ($\pm$ 0.0103)	&		0.3140	&			Rec\\
FCC 308	&	8.3831	($\pm$ 0.0012)&	-0.0076	($\pm$ 0.0013)&		0.1117	&			Int\\
FCC 312	&	8.6034	($\pm$ 0.0008)&	-0.2821	($\pm$ 0.0012)&		0.1101	&			Rec\\
FCC 179	&	8.787	($\pm$0.001)&	-0.0283	($\pm$0.0020)&		0.0548	&		Rec\\
		\hline
	\end{tabular}
\end{table*}

%Old version
%	Galaxy 	&	intercept		&	slope		&	RMSE		&	Mass &	Dist	&	 R$_{\rm e}$ (arcsec)	&	Infaller\\
%		\hline
%FCC119	&	8.452	($\pm$ 0.007)&	0.1555 ($\pm$ 0.0331)	&		0.1506	&	9.0	&	2.1	&	17.4		&		Int\\
%FCC090	&	8.3542	($\pm$ 0.0028)&	-0.0293	($\pm$ 0.0018)&		0.1179	&	8.9	&	1.7	&	12.1		&		Rec\\
%FCC113	&	8.065	($\pm$ 0.003)& 	0.005205	($\pm$ 0.003184)&		0.2328	&	8.3	&	1.21	&	20.56	&		Int\\
%FCC263	&	8.2166	($\pm$ 0.0013)&	-0.0208	($\pm$ 0.0015)&		0.1115	&	8.6	&	0.79	&	27.15	&		Rec\\
%FCC285	&	8.0658	($\pm$ 0.0030)&	-0.1362	($\pm$ 0.0039)&		0.3157	&	8.3	&	1.17	&	49.9		&		Rec	\\
%FCC290	&	8.824	($\pm$ 0.001)&	       -0.07732	($\pm$ 0.00264)&		0.1198	&	9.8	&	1.05	&	48.52	&		Int\\
%FCC306	&	7.863	($\pm$ 0.011)&	     0.04669 ($\pm$ 0.0103)	&		0.3140	&	7.47	&	1.69	&	9.7		&		Rec\\
%FCC308	&	8.3831	($\pm$ 0.0012)&	-0.0076	($\pm$ 0.0013)&		0.1117	&	8.6	&	1.69	&	37.11	&		Int\\
%FCC312	&	8.6034	($\pm$ 0.0008)&	-0.2821	($\pm$ 0.0012)&		0.1101	&	10.17 &	1.59	&	109.5	&		Rec\\
%FCC179	&	8.787	($\pm$0.001)&	-0.0283	($\pm$0.0020)&		0.0548	&	10.19 &	0.55	&	30.03	&		Rec\\

%
%\begin{figure}
%\includegraphics[width=0.4\textwidth]{HistF3D_Diff.pdf}
%    \caption{slopes difference}
%    \label{fig:GradDiff}
%\end{figure}

\subsection{TNG50 simulations from the IllustrisTNG Project}\label{sec:TNG50}

To further investigate the reliability of the suggested flattening from the observations, we perform a similar exercise using the high resolution TNG50 data  from the IllustrisTNG Project \citep{Pillepich19,Nelson19,Nelson19b}. First, we identify clusters with similar characteristics to Fornax cluster. The virial mass of Fornax is M$_{\rm vir}$ $\sim$ 3-6 $\times$ 10$^{13}$ M$_{\odot}$ from \citet{Drinkwater01}, however, for the sake of increasing our statistics, we expanded the search of Fornax-like clusters to friends-of-friends (FOF)  groups in the range of virial mass  $\sim$ 2-7$ \times$ 10$^{13}$ M$_{\odot}$. This gives a total  of 12 Fornax-like clusters in TNG50.

For these 12 Fornax-like clusters, we proceed to find simulated galaxies (or subhalos) with stellar masses similar to our Fornax-ELGs sample within each cluster's viral radius at z = 0. Although stellar mass is our primary discriminator, we also restricted our sample to SFRs within $\pm$0.2 dex range. This retrieves a total of 15 subhalos. Due to the resolution of TNG50, we are unable to find subhalos with stellar masses lower than $\sim$10$^{8.0}$ M$_{\sun}$. 

Once our sample is constrained, we estimated gas metallicity gradients as done in \citet{Hemler21} and \citet{Ma17}. First, the center of each galaxy is defined as the position of the particle (of any type) at the minimum potential, and we define the origin of the coordinate system in this position.  Gas cells with 
hydrogen number density $n_{\rm H}$ $<$ 0.13 cm$^{-3}$  are excluded to avoid contributions from diffuse gas outside the galactic disk. 
Following \citet{Ma17}, we do not fit the inner 1/4 part of the galaxy, since the gradient tends to be steeper or flatter in comparison with the outer star-forming region. 
Gas metallicities are estimated using the oxygen to hydrogen ratio 12+log($\rm{N_O}$/${\rm N_H}$).  These metallicities are systematically higher than the observational values we obtained following \citet{Dopita16}. However, the intrinsic differences between the TNG50 gradients should be indicative of any systematic difference.

\begin{figure*}
\includegraphics[width=0.83\textwidth]{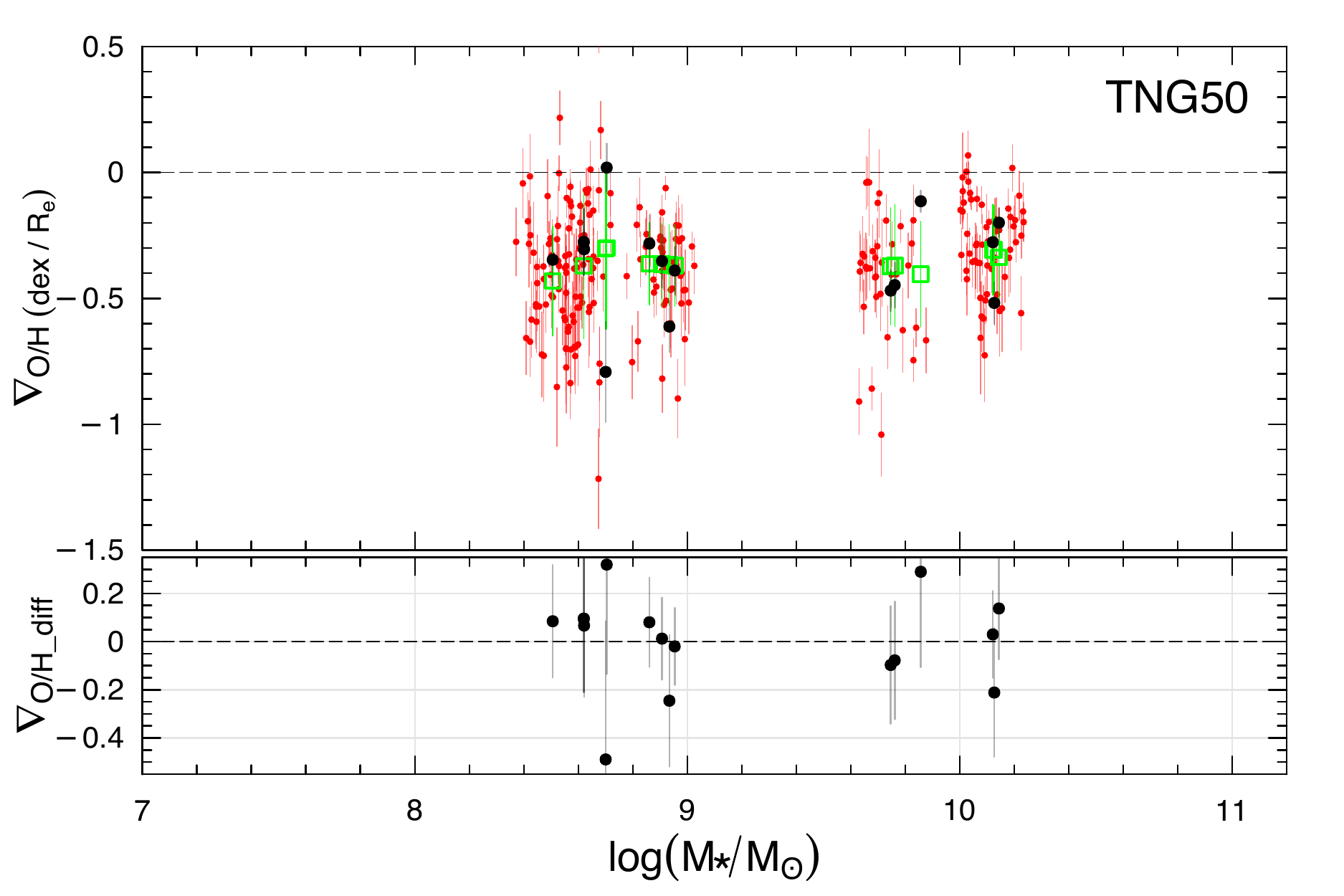}
    \caption{Gas metallicity gradients as a function of stellar mass for the subhalos  from the TNG50 simulation. The black circles correspond to the subhalos in Fornax-like clusters, while the red circles indicate subhalos  form our control sample (see text).  The green squares show the median value  $\tilde{\mathcal \nabla}$$_{\rm O/H}$ for the control sample, where the vertical line show their 1 $\sigma$ dispersion. The bottom panel shows the difference in metallicity gradient between each subhalo and the corresponding  $\tilde{\mathcal \nabla}$$_{\rm O/H}$ value for the control sample.}
    \label{fig:MetGradients_Illustris}
\end{figure*}

In contrast with   \citet{Hemler21} and \citet{Ma17}, and to be consistent with our observational data, our gas metallicity gradients are normalized to the R$_{\rm e}$ of each subhalo. An individual inspection to each gradient shows that fitting data out to 2 R$_{\rm e}$ results in reliable fits. Finally, only gradients with at least 16 gas cells are considered. This gives a final sample of 15 subhalos in Fornax-like clusters with reliable gradients. The galactocentric gradients are derived using  ``HYPERFIT", and the results are shown in Fig. \ref{fig:MetGradients_Illustris}.

\begin{figure*}
\includegraphics[width=0.4\textwidth]{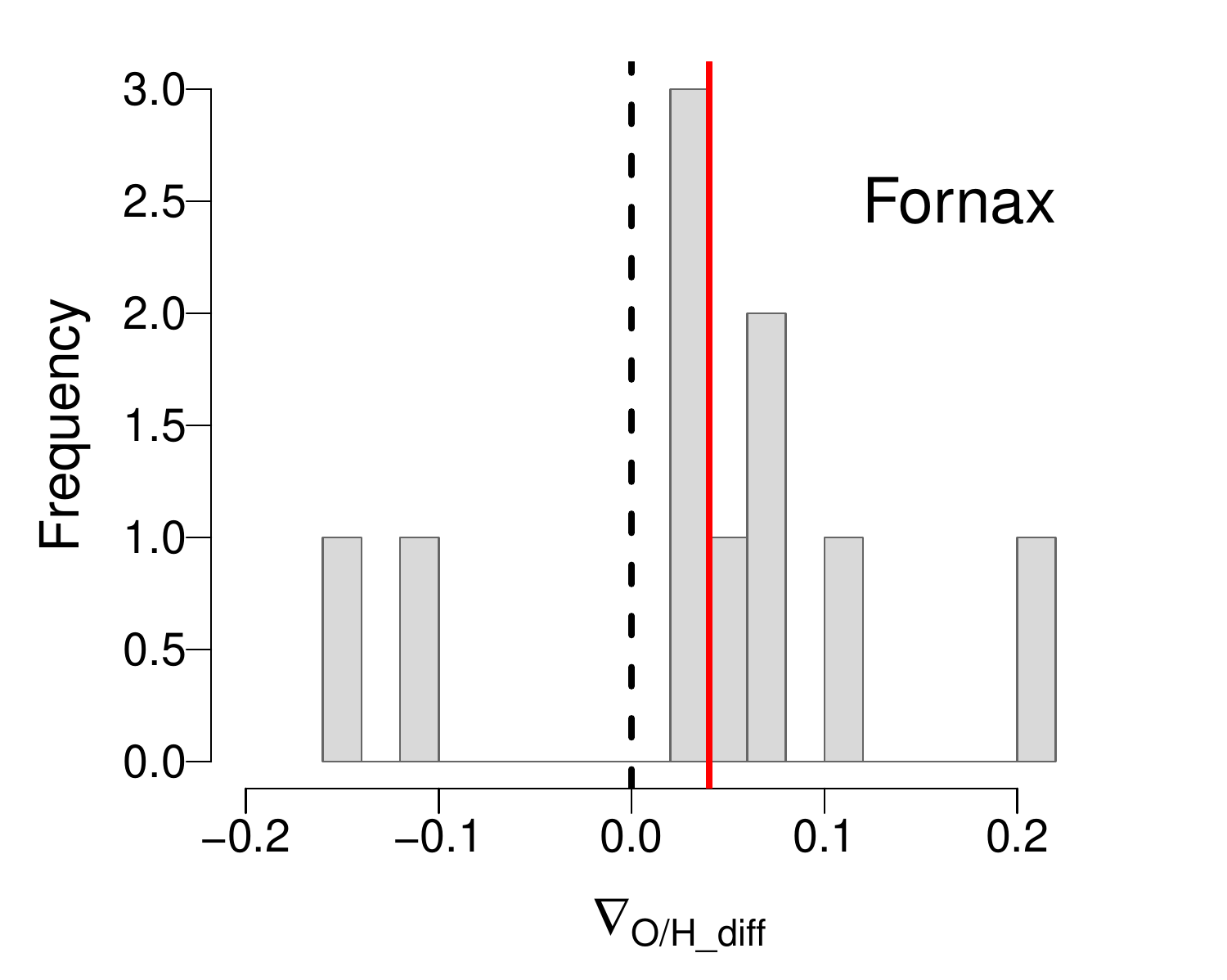}
\includegraphics[width=0.4\textwidth]{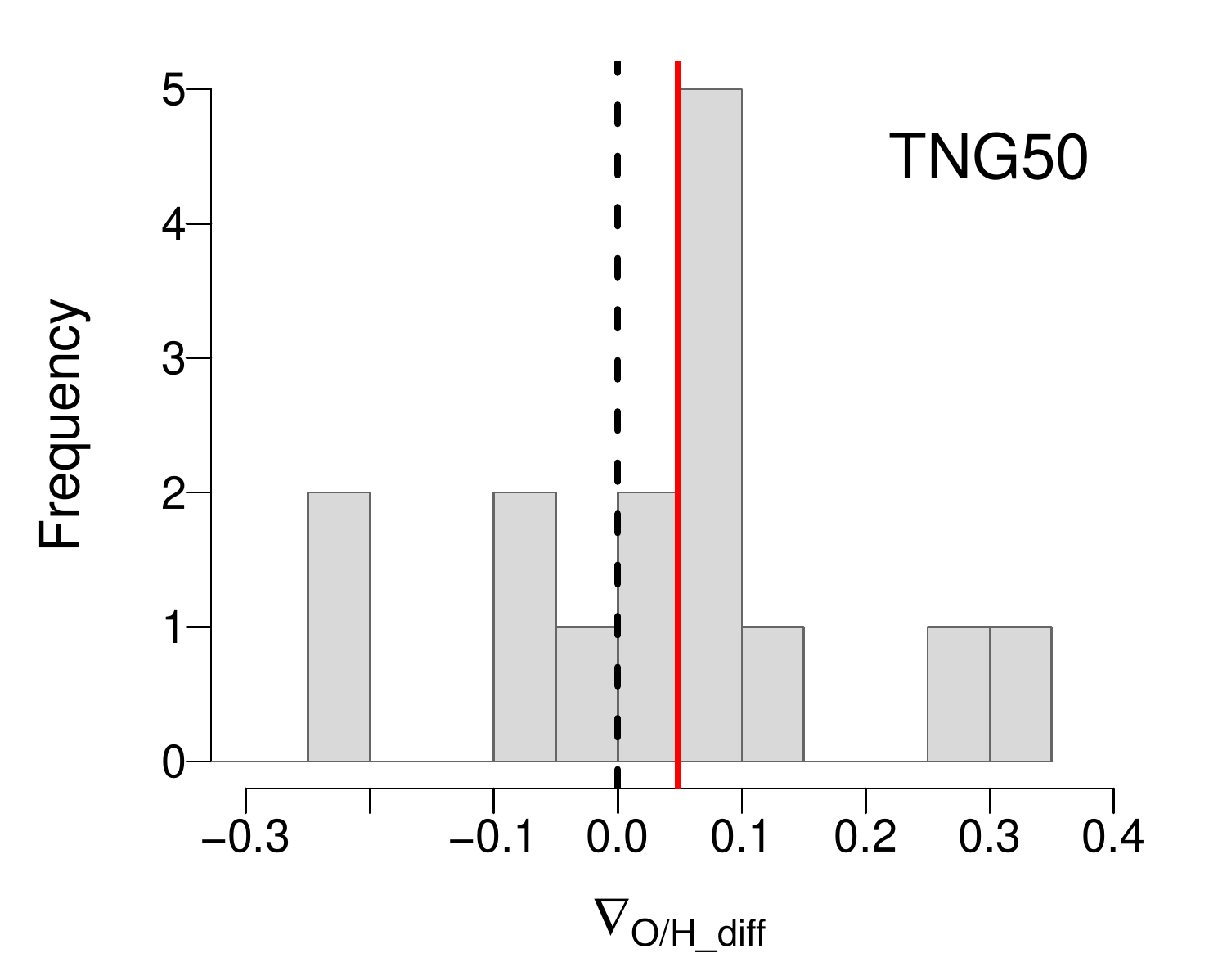}
    \caption{Left: Histogram of the differences between the gas metallicity gradients of Fornax ELGs. The red vertical line shows the median value of the differences ($\tilde{\mathcal \nabla}$$_{\rm O/H\_diff}$). Right: Same as the left panel but for the subhalos from TNG50.}
    \label{fig:GradDiff_F3DAndIllustris}
\end{figure*}

 We follow our observational methodology and define a control sample of subhalos in TNG50. We define the equivalent of 'field galaxies' in TNG50 as the
subhalos that are alone within a radius of 5$\times$R$_{200c}$, where R$_{200c}$ is the virial radius of the considered halo \citep[e.g., ][]{Mistani16}. In this way, we guarantee the central subhalos to be isolated from other central subhalos.
%La muestra de control que tomé se obtuvo considerando los subhaloes centrales que no poseen ningun otro subhalo central dentro de un radio de 5R_200c, con R_200c el radio virial del halo al que pertenece el subhalo. Recomiendo citar Mistani et al. 2016, en el que se explica porqué la evolución de los centrales difiere de la de los satelite subs.
%the definition of \citet{Joshi20}, who define the equivalent of 'field galaxy' in illustris as central subhalos in a cluster that have not been accreted by any other subhalo, and consequently have always been central up to z = 0. 
The control subhalos are then mass-matched, within a $\pm$0.15 dex range, to each one of the 15 Fornax-like subhalos described previously. The measurements of the gas metallicity gradients is performed as previously described, giving us a final control sample  of 252 subhalos. The resulting gradients for  Fornax-like and control subhalos are shown in Fig. \ref{fig:MetGradients_Illustris}.  Similarly to our observational methodology, we estimated the median value   $\tilde{\mathcal \nabla}$$_{\rm O/H}$  for the control sample of each Fornax-like subhalo. The  median values are shown as green squares in the upper panel of Fig. \ref{fig:MetGradients_Illustris}, and the differences between the Fornax-like and control $\tilde{\mathcal \nabla}$$_{\rm O/H}$  are shown in the lower panel of the same Figure. The histogram of these differences is displayed in Fig. \ref{fig:GradDiff_F3DAndIllustris}, showing a $\sim$0.05 dex/R$_{\rm e}$  difference towards more positive gradients for Fornax-like subhalos. Even though the difference is small, it is consistent and in the same direction as we find for our observational sample.

Different physical phenomena or mechanisms could explain a flattening in the gas metallicity gradient of cluster galaxies. For instance, minor mergers of chemically poor galaxies could potentially flatten the gradient of the most massive galaxy. 
 Even though mergers are more likely to happen before the cluster is virialized, in order to assess or discard the possible role played by minor and major mergers in Fornax-like clusters, we analyze the merger history of our TNG50 sample.  To this aim, we use the merger tree history estimated by  \citet{Rodriguez15} for each subhalo in our Fornax-like and control sample.  Major mergers are defined to occur for a mass ratio higher than 1:4 between the merging subhalos, while minor mergers for a mass ratio in between  1:10 and 1:4. We considered the number of major and minor mergers of each subhalo up to redshift  2. 
%In Fig. \ref{fig:GradDiff_Mergers} we can appreciate no evident trend between the metallicity gradient and the number of mergers. 

\begin{figure*}
\includegraphics[width=0.5\textwidth]{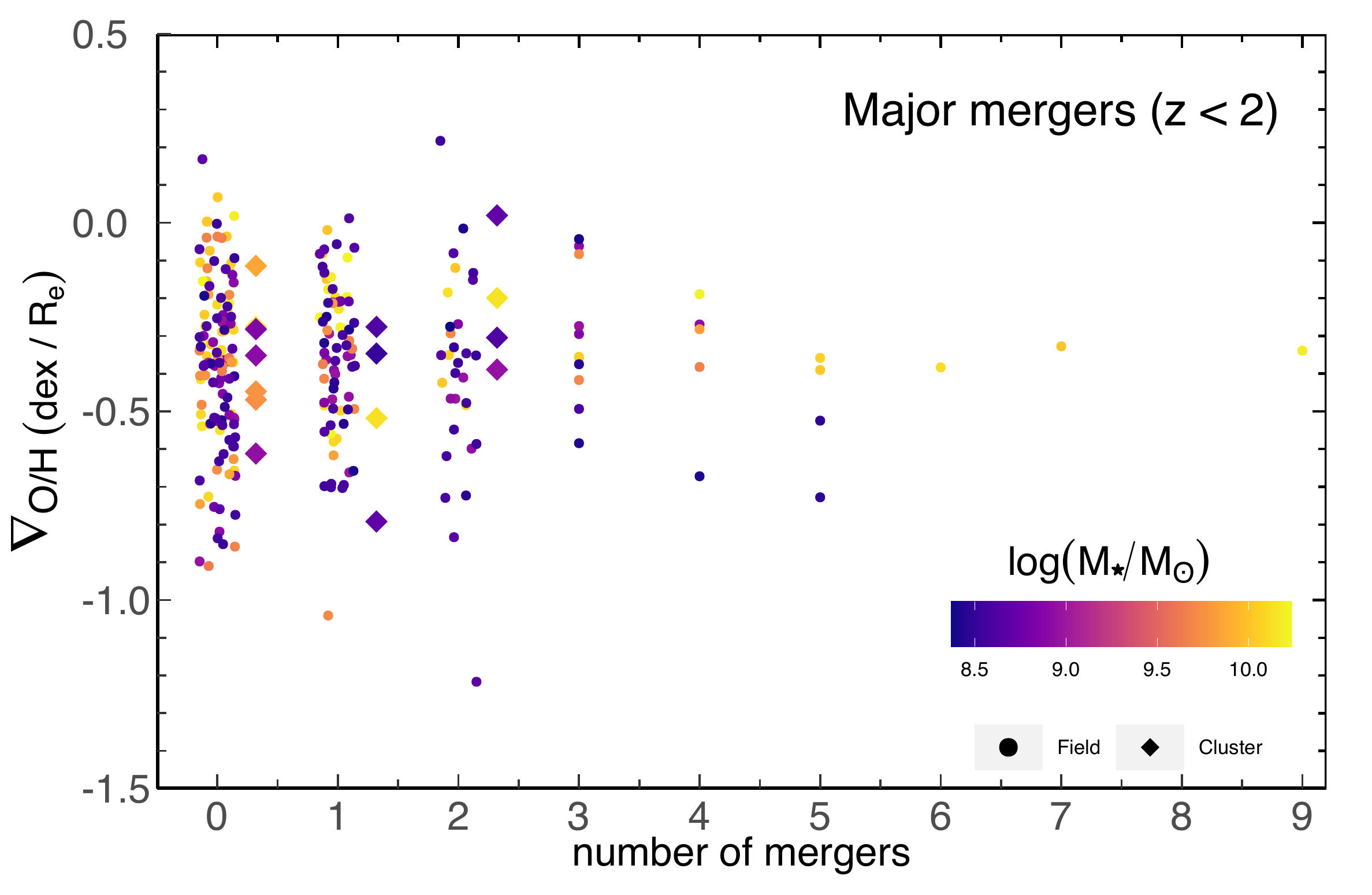}
\includegraphics[width=0.5\textwidth]{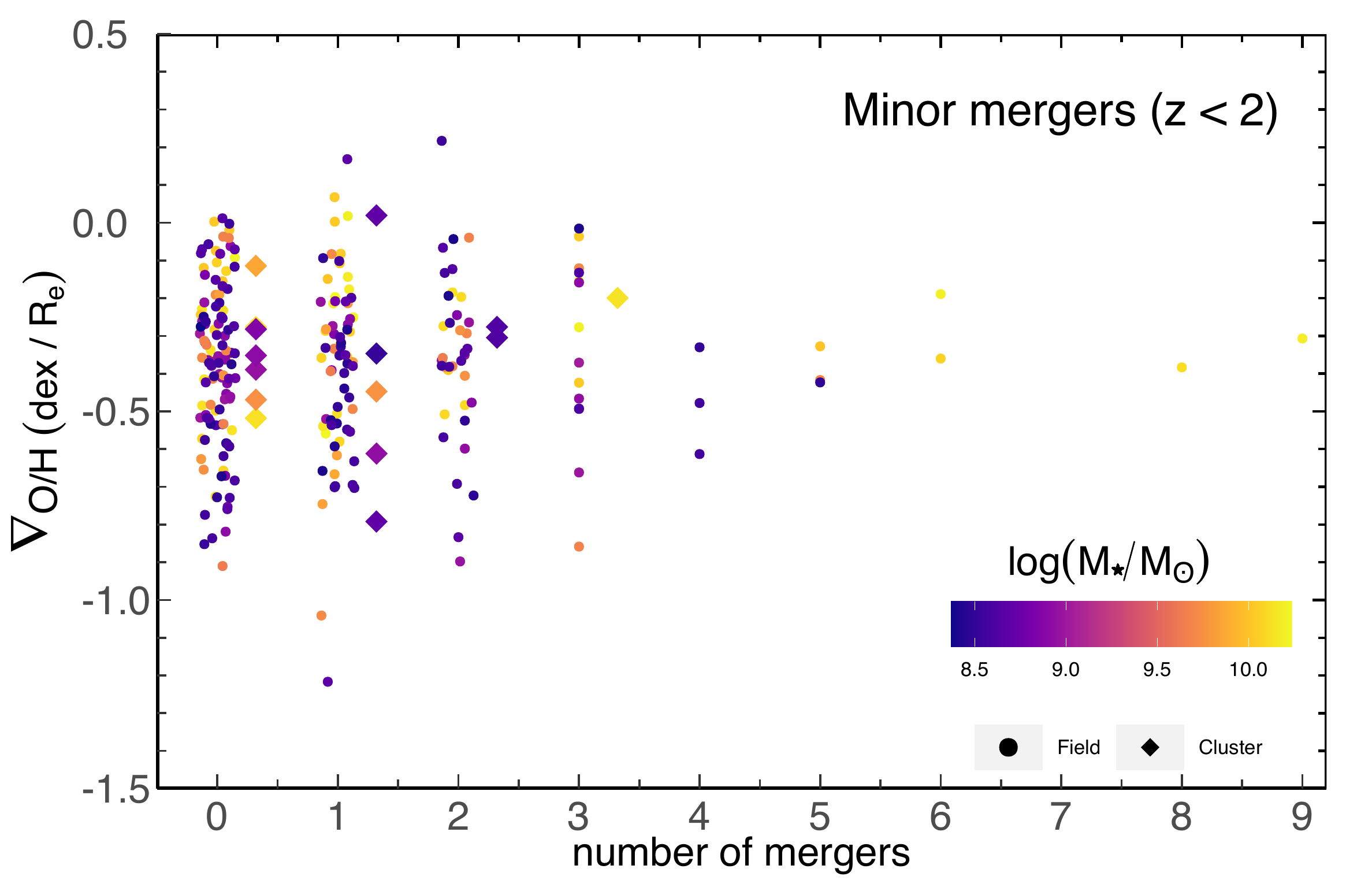}
    \caption{Gas metallicity gradient as a function of number of mergers up to redshift $<$ 2. Left and right panel indicate Major and Minor mergers, respectively. Circles indicate subhalos from the control sample, while diamonds the subhalos in Fornax-like clusters. All the symbols are color coded as a function of their stellar mass.  For the sake of clarity, all the subhalos in the control sample with a number of mergers between 0 and 2 were assigned random shifts (in the range $\pm$0.25) around their actual value, while the subhalos in Fornax-like clusters are shifted to the right.}
    \label{fig:GradDiff_Mergers}
\end{figure*}

If recent mergers have had an influence in flattening the gas metallicity gradient, we would expect a relation between the gradient and the number of mergers for galaxies in clusters. However,  in Fig. \ref{fig:GradDiff_Mergers} we can not see any evident trend for either minor or major mergers. The same result is found when we limit to  different mass bins, and when we constrain the merger history to a more recent past (e.g., z $\sim$ 1).  Hence,  we can conclude that the merger history in Fornax-like cluster galaxies is not directly responsible for the flattening of the gas metallicity gradients. 

 Finally, we investigate the role played by ram pressure stripping (RPS) produced by the ICM. We follow \citet{Yun19}  and estimate the (mode of the) ram pressure acting on the galaxy based on the equation of  \citet{Gunn72}: $P_{\rm ram} =$ Mode ($\rho_{\rm medium} \times v_{\rm rel}^2$ ),  where  $\rho_{\rm medium}$ is  the density of the medium gas cells, and $v_{\rm rel}$ is the relative velocity of the gaseous component of a satellite with respect to its surrounding medium. 
The relative velocity is defined as $v_{\rm rel}$ = Mode($v_{\rm satellite \ gas}$ - $V_{\rm medium}$), where $v_{\rm satellite \  gas}$  is the velocity of the gas cells bounded to the satellite galaxy, and $V_{\rm medium}$ is the typical velocity of the medium gas cells.
Each medium gas cell is constrained to gas cells bound to the central subhalo within 20 times the stellar half mas radius around it.

\begin{figure}
\includegraphics[width=0.51\textwidth]{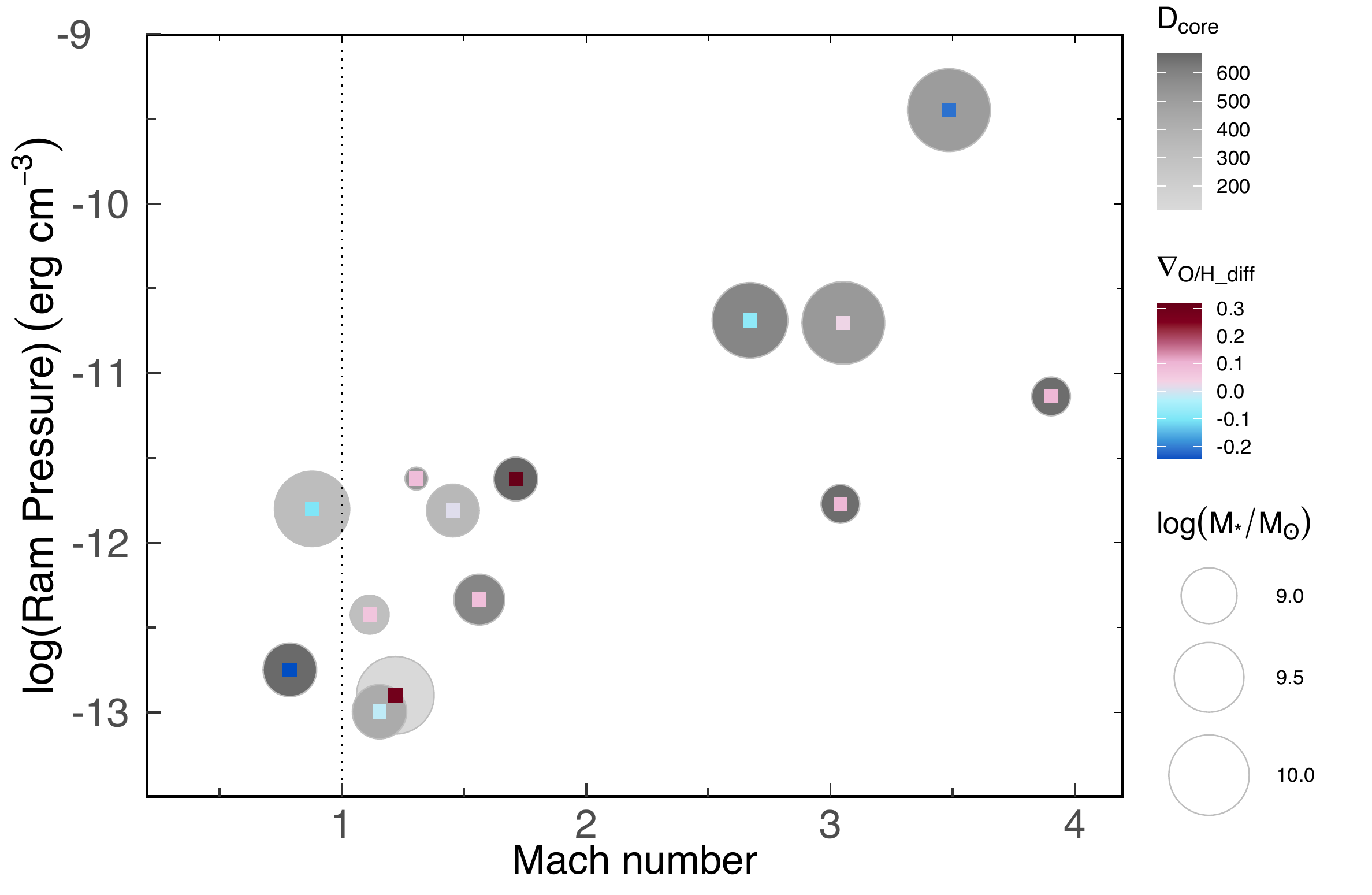}
    \caption{ Ram pressure as a function of Mach number  for the TNG50 subhalos in Fornax-like clusters. The small squares are color coded according to ${\mathcal \nabla}$$_{\rm O/H\_diff}$, while the grey circles are color coded according to the cluster-core distance of each subhalo. The circle size corresponds to the stellar mass of each subhalo as indicated in the right bar. The vertical dotted line indicates the threshold between supersonic ($\mathcal{M}$ $>$ 1) and subsonic ($\mathcal{M}$ $<$ 1) velocities. }
    \label{fig:RPS}
\end{figure}

We also  estimated the Mach number ($\mathcal{M}$), defined as the speed at which an object moves relative to a fluid divided by the sound speed of the fluid: $\mathcal{M}$ = $v_{\rm rel}$/ Mode($c_{s, medium}$). We are able to calculate theis parameter for 14 subhalos in our Fornax-like sample. Figure \ref{fig:RPS}  shows the Mach number vs ram pressure estimates, color coded by ${\mathcal \nabla}$$_{\rm O/H\_diff}$ and cluster-core distance, while the circle sizes correspond to the stellar mass of each subhalo. The distribution of our data show a general agreement with those of \citet{Yun19}, where higher Mach numbers show larger ram pressures. A further interpretation of this figure is given in the \S \ref{sec:Discussion}.

%In regards of this, as in Joshi, they defined a 'field galaxy' as a CENTRAL SUB (central subhalo in a cluster) that hasn't been accreted by any other sub, and consequently has been central FOREVER

%Reff how it is estimated

%of the galactic star-forming region because of its proximity to the galactic central region, which often possess a gradient that is either much steeper or much atter than that of the star-forming region.

\begin{figure*}
\includegraphics[width=0.83\textwidth]{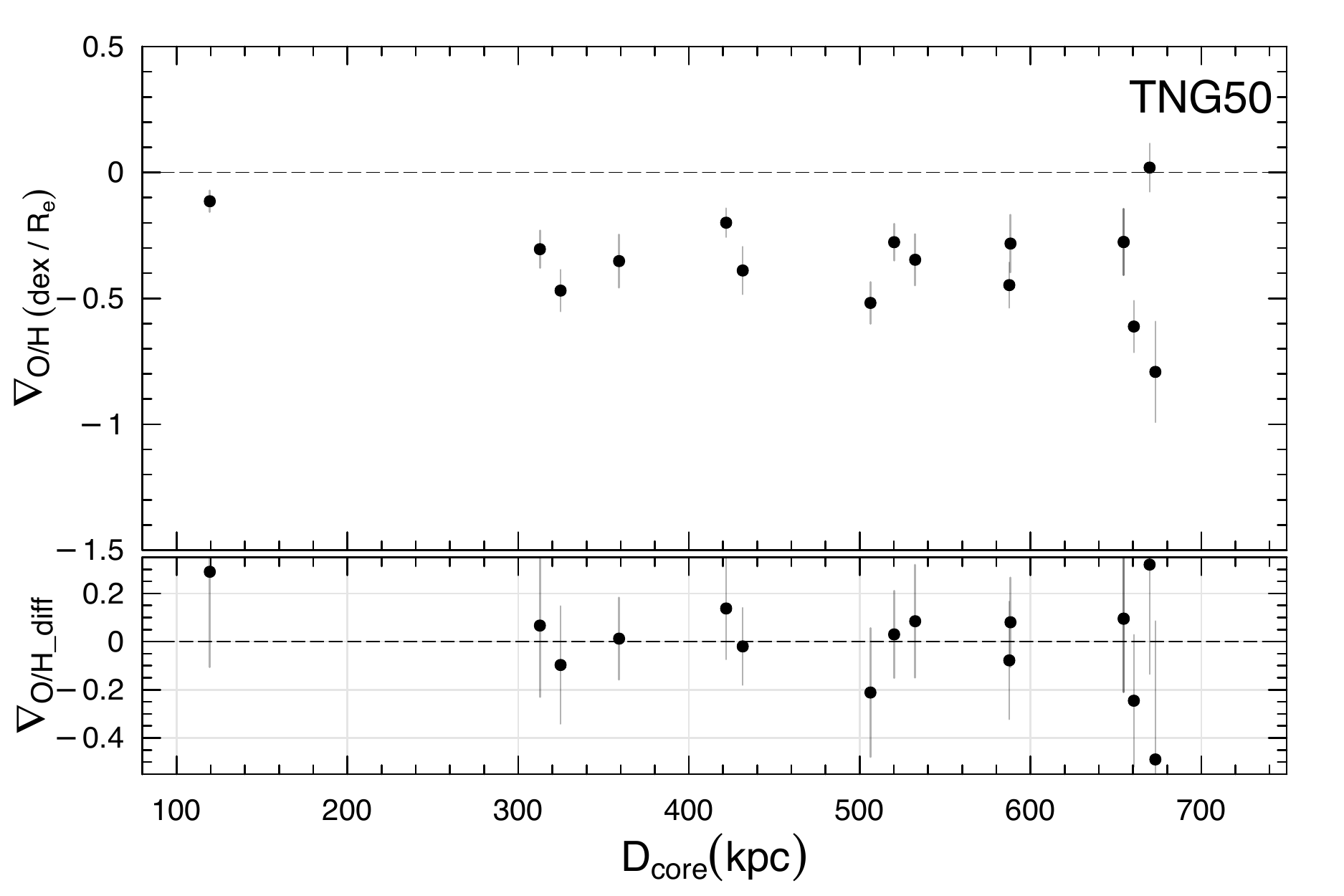}
    \caption{Gas metallicity gradients as a function of distance to cluster core for the subhalos  from  TNG50. The black circles correspond to the subhalos in Fornax-like clusters. The bottom panel shows the difference in metallicity gradient between each subhalo and the corresponding   $\tilde{\mathcal \nabla}$$_{\rm O/H}$ value for the control sample.}
    \label{fig:MetGradients_Dist_Illustris}
\end{figure*}

\section{Discussion}\label{sec:Discussion}

Late-type galaxies show, in general, negative metallicity gradients, likely as a result of an inside-out formation scenario \citep[e.g., ][]{White91,Perez13}, where stars have been forming in their centres for longer, building up the metallicity there. Variations in the gas metallicity gradient can potentially indicate feedback processes due to cluster environment, either due to inflows/outflows of pristine or enriched gas that can alter the metallicity gradient, mergers, or tidal and flyby interactions, among others. However, the observational evidence of the effect of cluster environment on gas metallicity gradients  is rather limited so far.

In this paper we analyzed the gas metallicity gradients of ELGs in the Fornax cluster. Our gas metallicity gradients follow the general trend with stellar mass previously reported by \citet{Belfiore17} and \citet{Poetrodjojo21}, see Fig. \ref{fig:MetGradientsWithControl}. Since our sample of galaxies span a stellar mass range between 10$^{7.4}$ and 10$^{10.2}$ M$_{\odot}$, we are unable to comment about the observed flattening in the gradients for more massive galaxies.  

By comparing Fornax-ELGs gradients with those of galaxies from a control sample, we find that eigth of our ELGs galaxies show  more positive gradients, and a general  median difference of $\sim$0.04  dex/R$_{\rm e}$  for the whole sample. We did a similar exercise with data from the TNG50 simulation by selecting subhalos in Fornax-like clusters and comparing them with a sample of control subhalos. TNG50 data indicate  a flattening ($\sim$ 0.05  dex/R$_{\rm e}$ ) for subhalos  in Fornax-like clusters. 

Our findings agree with previous results that report a flattening in the metallicity gradients in galaxy clusters \citep{Franchetto21}, and in interacting galaxy pairs \citep{Kewley10,Rupke10b,Rupke10a}. Since galaxy interactions produce a flattening in the gas metallicity gradients, it is natural to expect this result in the very early stages of merger systems. However, according to simulations, the number of encounters within 10 kpc is about 10 per galaxy in the Hubble time. Only about 0.1\% of such encounters are expected to result in a merger. This gives a merger probability of 22–29\% per halo \citep{Gnedin03}. Altough most mergers take place before the cluster virializes.

Our sample of ELGs is within the virial radius of the Fornax cluster, and hence the  incidence of mergers is expected to be low \citep[see also][]{Joshi20}. In Sec. \ref{sec:TNG50}  we analyzed the merger history for our sample of Fornax-like subhalos, and indeed found that in comparison with the control subhalos, the incidence of  minor and major mergers is significantly lower up to z$\sim$2. Hence, it is highly unlikely that mergers are responsible for the observed flattening of the gas metallicity gradients.

Another explanation proposed by  \citet{Franchetto21} indicates that the cluster galaxies that show flatter gradients might have fallen into the cluster sooner and hence experienced environmental effects for longer time. All the galaxies in our sample are infalling into the cluster, and were classified as recent and intermediate infallers by \citet{Iodice19}. As indicated in Fig. \ref{fig:MetGradientsWithControl}, eight of our ELGs, four intermediate and four recent infallers, show flatter gradients with respect to the control sample, whereas the remaining two galaxies, recent infallers, show steeper gradients. On the other hand, all our four intermediate infallers show more positive gradients.

%Semianalytic models of galactic encounters (Merritt 1983) predict the segregation of galaxies by mass and the formation of a central cD galaxy. 
% The giant cD grows by swallowing other massive galaxies in a process that widens the gap between the first and the second ranking galaxies (Ostriker \& Hausman 1977). Mass segregation in clusters is also supported by numerical simulations (Frenk et al. 1996). Recently, Dubinski (1998) confirmed that the central galaxy, rising by mergers of its neighbors, follows a de Vaucouleurs profile and would be classified as a giant elliptical galaxy.

%https://iopscience.iop.org/article/10.1086/344636/fulltext/50024.text.html
 %    Massive galaxies are strongly clustered at the end of the simulations. Together with the remaining substructure, this significantly increases the probability of galactic collisions. 

%The tidal perturbation induced by flybying neighbors may create galactic bars

Tidal features (e. g., tail, arms or streams) have been considered as possible proxies of merger systems \citep[e.g., ][]{Oh08,Hood18}. Even though galaxy mergers have been  the focus of several investigations due to their ability to re shape galaxy properties such as morphology, the effect of smaller bodies such as orbiting satellites can produce observable perturbations too. In addition, tidal interactions are also responsible of the observed HI deficiency observed in several cluster galaxies \citep[e.g., ][]{Gnedin03,Hughes13}. 

In this paper we examined two possible processes that could potentially cause a flattening in the gas metallicity gradients, flybys and ram pressure stripping.
 Flybys are rapid and transient events that occur when two independent galaxy halos interpenetrate but detach at a later time \citep[e.g., ][]{Sinha12, An19}. The importance of flybys in galaxy evolution has been recently acknowledged since multiple interactions with two or more neighbours are on average flyby-dominated. According to \citet{An19}, flybys substantially outnumber mergers (by a factor of five) toward z = 0. Hence the  contribution of flyby’s to galactic evolution is stronger than thought.
%Flybys can create galactic bars, ring structures, change the direction of the galaxy’s angular momentum, and enhanced or quenched star formation, among others \citep[for a review see][]{An19}. 
An extreme example of frequent high speed galaxy encounters is known as galaxy harassment, and it is a likely mechanism for inducing a morphological transformation  \citep{Moore96}. 
 Another possible process at play is RPS, in this scenario, RPS can create a lopsided pressure on the disc of gas in the galaxy, causing the lower metallicity gas at the outskirts to lose angular momentum and fall towards the centre of the system, flattening the observed gradient.  Whether the galaxy-galaxy (i.e., flybys) or the cluster-galaxy (i.e., RPS) interactions dominate is yet to be understood.

 To elucidate the impact these processes have on galaxies, we analysed the Mach number and ram pressure in our sample of Fornax-like subhalos in TNG50 (Fig. \ref{fig:RPS}). According to \citet{Yun19}, galaxies at supersonic velocities ($\mathcal{M}$ $>$ 1), will produce discontinuous features in the fluid such as shocks and contact discontinuities, whereas  subsonic motions ($\mathcal{M}$ $<$ 1) would allow smooth changes.  Our sample shows negative ${\mathcal \nabla}$$_{\rm O/H\_diff}$ for suhbalos with $\mathcal{M}$ $<$ 1, whereas those in the range 1 $<$ $\mathcal{M}$ $<$ 2, show positive ${\mathcal \nabla}$$_{\rm O/H\_diff}$, corresponding to flatter (or more positive) gradients with respect to their control sample. This indicates that supersonic velocities might be the main cause in producing flatter/positive metallicity gradients. As for subhalos with higher Mach numbers ($\mathcal{M}$ $>$ 2),  \citet{Yun19} suggest that they might in fact be flybys. Our TNG50 sample shows five possible flybys, mostly located at large cluster-centric distances. However, their gas metallicity gradients, show either positive or negative  ${\mathcal \nabla}$$_{\rm O/H\_diff}$. Therefore, our sample suggests that while flybys could produce flatter gradients, subhalos at supersonic velocities with 1 $<$ $\mathcal{M}$ $<$ 2, are likely to have flatter/more positive gas metallicity gradients.

Back to our observational sample, with the exception of  FCC 179 and FCC 290, the rest of the galaxies in our sample show a disturbed morphology. It is difficult to address which kind of interactions have played a major role in defining it. Although in some specific cases the photometric images from \citet{Raj19} provide some hints. For instance, a flyby might have happened in FCC308 and FCC312, where a thick disk is observed, while FCC 312 shows an extended warped tail.  In addition, the gas metallicity gradient of FCC 290 (see Fig. \ref{fig:PanelsMetGradientsWithControl}) suggest an inflection at 0.74 R/R$_{\rm e}$ to a flatter/positive gradient that corresponds to a drop in the  H$_2$-to-dust ratio. As indicated by  \citet{Zabel21}, this drop means an outside-in stripping of the gas/dust disc, which results in  a lower integrated H2-to-dust ratio. This is an example on how stripping of gas can produce flatter/positive gradients.

From our sample, FCC 308, FCC 263 and FCC 090 have been identified to have disturbed CO by \citet{Zabel19}. In addition, Fornax dwarf galaxies have been shown to have a molecular gas fraction about an order of magnitude smaller than expected, while several galaxies also show asymmetric gas distributions \citep{Zabel19}.  It is uncertain whether these effects are due to ram pressure stripping or flyby interactions,  but in both cases, disturbing the gas   could result in flatter gas metallicity gradients. In fact, simulations show that strong perturbations can stir the gas and drive galactic-scale motion in the ISM, causing gas/metal re distribution on galactic scales \citep{Ma17}.

Regarding the general properties of our sample, the integrated metallicity of our Fornax sample shows slightly higher metallicities ($\sim$0.045 dex) in comparison with a control sample from the SDSS survey. Our result is in agreement with previous works that report slightly higher ($\sim$0.05 dex) metallicities for late-type galaxies in clusters \citep{Ellison09,Scudder12,Gupta16,Coenda20}.  In addition, our sample of 10 ELGs is statistically small, and hence a larger sample is needed to properly quantify any difference in the integrated gas metallicity of Fornax-like clusters.

%This is a good paper to get citations!
%In agreement with our results, \citet{Franchetto21} found recently flatter metallicity gradients for cluster galaxies using data from the GASP \citep{Poggianti17} and MANGA \citep{Bundy15} surveys...

On the other hand, we detect signs of a mass and metallicity segregation, as described in Sec. \ref{sec:SampleCharacterization}. This is in agreement with models and numerical simulations that predict the segregation of galaxies by mass and the formation of a central BCG galaxy \citep[e.g.,][]{DeLucia07}.
Our results agree with \citet{Gupta16, Gupta17}, who observed a negative metallicity gradient with clustercentric distance for part of their cluster sample. As indicated by  \citet{Kim20}, the mass of the whole cluster may play an important role defining the timespan of dynamical friction, and therefore mass segregation may be detected only in low mass clusters.

%Explanation: As a massive objects M moves through a sea of particles, the particles passing by are accelerated towards the object, As a result, the particle number density behind the object is higher than that in front of it, and the net effect is a drag force, or dynamical friction, on the object.

Finally, a source of uncertainty in our study is related to the different spatial resolutions of the surveys we are using. While SAMI has a spatial resolution of 2.16 arcseconds, galaxies observed with MUSE have a spatial resolution of 0.2 arcseconds. According to \citet{Acharyya20}, the metallicity gradients are sistematically shallower than the true value for surveys with a lower spatial resolution. This indicates that the true metallicity gradients for the SAMI sample could be more steep, and hence the difference with our Fornax sample could be even larger.

A second caveat relates to the projected distance to the BCG shown in Fig. \ref{fig:MetGradientsWithDistance}. Altough the distances could change with a more precise distance measurements such as Planetary Nebulae indicators \citep[e.g., ][ currently available only for some early-type galaxies in the Fornax cluster]{Spriggs20,Spriggs21},  it is uncertain whether or not a pattern would emerge based on the current results.

\section{Conclusions}\label{sec:Conclusions}

As part of the Fornax-3D project, in this paper we analyze the gas metallicity gradients of 10 ELGs in the Fornax cluster observed with MUSE. We used a control sample formed by galaxies from B19 and the MAD and SAMI surveys. Gas metallicities were estimated consistently for Fornax and control galaxies. 
A summary of our findings is listed below:

   \begin{itemize}
   
         \item The integrated gas metallicity of Fornax ELGs follows the general M-Z relation for SDSS galaxies.  In agreement with previous results, our Fornax sample show slightly higher metallicities by 0.045 dex in comparison with the control sample.

      \item Galaxies in the Fornax cluster show signs of  mass and metallicity segregation, likely as the result of dynamical friction having a stronger effect on medium mass clusters.
      
      \item Our data suggest ELGs in the Fornax cluster exhibit flatter (or more positive) metallicity gradients by $\sim$0.04  dex/R$_{\rm e}$  in comparison with our control sample.

      \item The derived metallicity gradients for Fornax galaxies as a function of stellar mass follows the general trend reported in previous works. That is, flatter gradients for low mass galaxies that steepen for more massive galaxies, and then flattens slightly again for more massive galaxies \citep[e.g.,][]{Belfiore17, Poetrodjojo21}.
      
           \item   Our data suggest there is no relation between the flattening of the metallicity gradients and the projected distance to the BCG.
      
           \item  All our Fornax ELGs classified as intermediate infallers (although there are only four) show more positive metallicities gradients with respect to the control sample.

%           \item {\bf No  pattern is found between the metallicity gradient of recent and intermediate infallers when analyzed as a function of stellar mass, and projected distance to the BCG. }

      \item We identified 12 Fornax-like clusters in the  TNG50 simulations, and selected a sample of 15 subhalos mass matched with our Fornax sample. Similar to the observations, we identified a flattening of $\sim$0.05 dex/R$_{\rm e}$  for subhalos in Fornax-like clusters. Additionally, no relation is found between the flattening and their distance to the cluster core.  
      
%      We identified a sample of 15 subhalos in Fornax like clusters in the Illustris TNG50 simulation. By comparing their metallicity gradients with a control sample, we find a sistematic flattening of $\sim$0.04  dex/R${\rm e}$. 

%The observed flattening can be a result of different environmental effects, such as flybys or ram pressure stripping. 

 \item  We estimated the Mach number and ram pressure for our sample of  TNG50 subhalos. The data indicate flatter or more positive gradients for subhalos with Mach numbers in the range 1 $<$ $\mathcal{M}$ $<$ 2, likely due to supersonic velocities that also scales with larger RPS values. On the other hand,  subhalos with $\mathcal{M}$ $>$ 2 are likely to be flybys and show either positive or negative   ${\mathcal \nabla}$$_{\rm O/H\_diff}$. 
 
  \item  Most of the ELGs in our MUSE sample show disturbed morphologies, hence, as the simulations suggest, the observed flattening could be the result of a combination of cluster-galaxy interaction (such as ram pressure stripping), or galaxy-galaxy interactions (such as flybys).

   \end{itemize}

\begin{acknowledgements}
We acknowledge the careful examination by the referee. 
We thank Mark den Brok and Jarle Brinchmann for providing us extra information for the MAD galaxies. 
We thank Henry Poetrodjojo for kindly providing us the SAMI data needed for this work.
GvdV acknowledges funding from the European Research Council (ERC) under the European Union's Horizon 2020 research and innovation programme under grant agreement No 724857 (Consolidator Grant ArcheoDyn). E.M.C. acknowledges support by Padua University grants DOR1935272/19, DOR2013080/20, and DOR2021 and by MIUR grant PRIN 2017 20173ML3WW-001. 
RMcD is the recipient of an Australian Research Council Future Fellowship (project number FT150100333).  This research was supported by the Australian Research Council Centre of Excellence for All Sky Astrophysics in 3 Dimensions (ASTRO 3D), through project number CE170100013.

\end{acknowledgements}

   \bibliographystyle{aa} % style aa.bst
   \bibliography{F3D_Gradients_V2.bib} % your references Yourfile.bib

%
%

\begin{appendix} %First appendix
\section{Maps}
In this section we present the \Ha\ flux, BPT class, and gas metallicity maps for each one of the galaxies analyzed in this paper.\\ \vspace{0cm}

\begin{figure}[h!] 
	\includegraphics[width=1.0\textwidth]{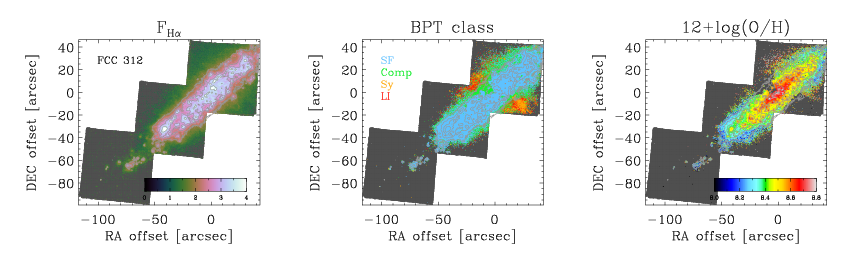}
	\includegraphics[width=1.0\textwidth]{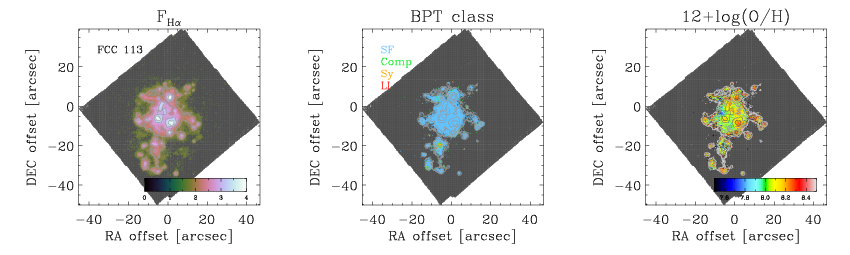}
	\includegraphics[width=1.0\textwidth]{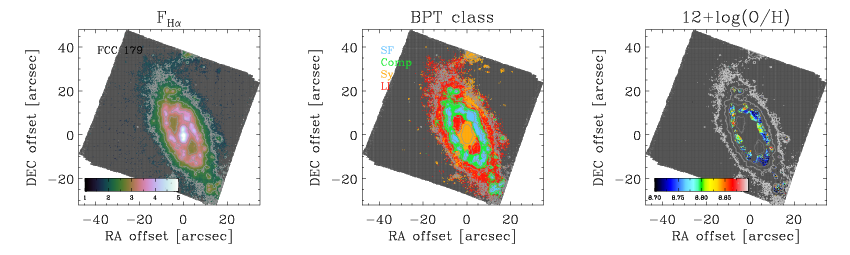}
	\caption{From left to right, \Ha\  flux, BPT classification, and gas metallicity maps for FCC 312, FCC 113, and FCC 179.  The inset color bars show the \Ha\ flux (left panel), gas metallicity (right panel), and the BPT classification (star forming, composite, Seyfert, and LINER, middle panel)}. 
        \label{fig:MapsAndBPT}
\end{figure}

\begin{figure*}
	\includegraphics[width=1.0\textwidth]{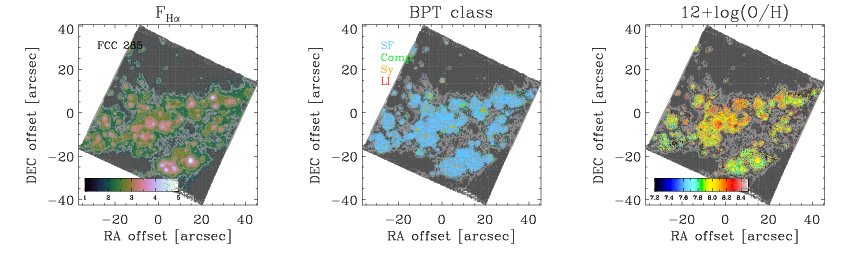}
	\includegraphics[width=1.0\textwidth]{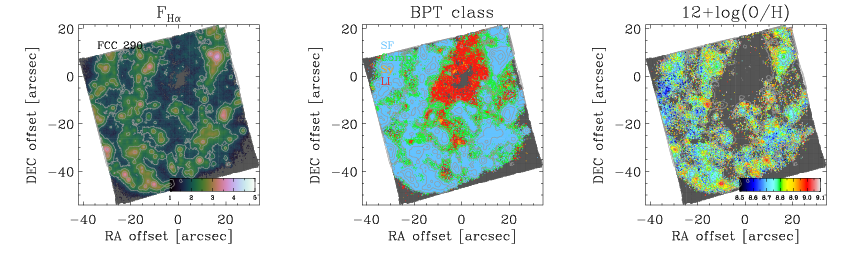}
	\includegraphics[width=1.0\textwidth]{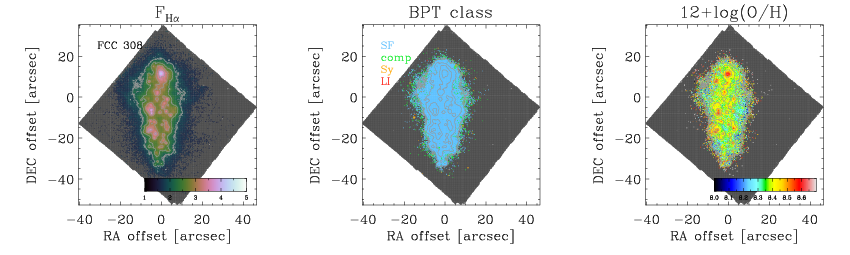}
	\includegraphics[width=1.0\textwidth]{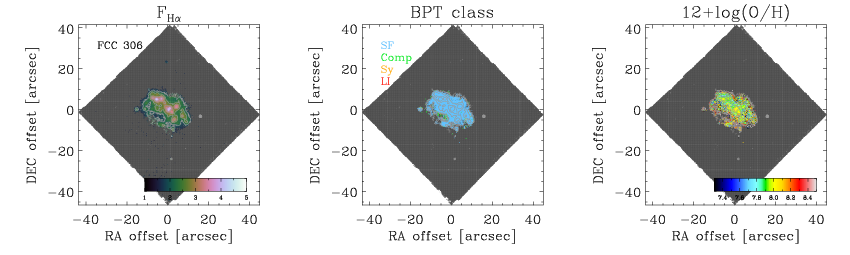}
	\caption{Similar to Fig. \ref{fig:MapsAndBPT} for FCC 285, FCC 290, FCC 308, and FCC 306.}
        \label{fig:MapsAndBPTb}
\end{figure*}

\begin{figure*}
	\includegraphics[width=1.0\textwidth]{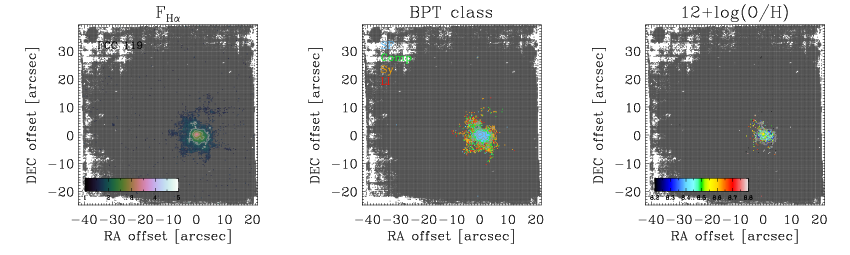}
	\includegraphics[width=1.0\textwidth]{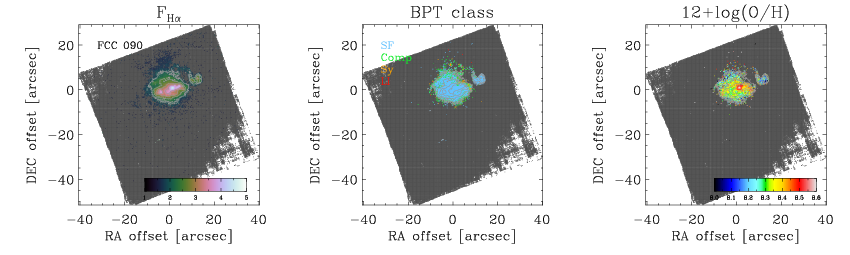}
	\includegraphics[width=1.0\textwidth]{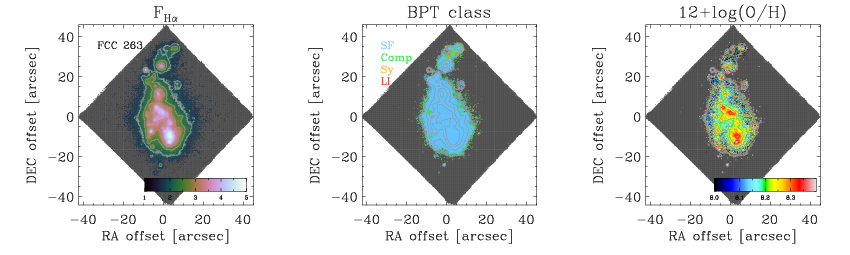}
	\caption{Similar to Fig. \ref{fig:MapsAndBPT} for FCC 119, FCC 090, and FCC 263.}
        \label{fig:MapsAndBPTc}
\end{figure*}

\end{appendix}

\end{document}